\documentclass[12pt,superscriptaddress,nature]{revtex4-2}

\usepackage[utf8]{inputenc}
\usepackage{graphicx}
\usepackage[usenames,dvipsnames]{xcolor}
\usepackage{epstopdf}
\usepackage{amsmath,amsthm,amssymb}
\usepackage{latexsym}
\usepackage{bm}
\usepackage{dcolumn}
\usepackage{braket}
\usepackage[normalem]{ulem}
\usepackage{times}
\usepackage{hyperref}
\usepackage[section]{placeins}

\newcommand{\FeTeSe}{Fe$_{1+y}$Te$_{1-x}$Se$_{x}$}
\newcommand{\FeTeS}{Fe$_{1+y}$Te$_{1-x}$S$_{x}$}

\newcommand{\br}{\mbox{\boldmath$r$}}

\newcommand{\bQ}{\mbox{\boldmath$Q$}}
\newcommand{\bP}{\mbox{\boldmath$P$}}
\newcommand{\bM}{\mbox{\boldmath$M$}}
\newcommand{\bsigma}{\mbox{\boldmath$\sigma$}}

\begin{document}
\bibliographystyle{naturemag}

\title{Multicritical superconductivity in iron chalcogenide}
\title{Magnetism, superconductivity, and surface state near criticality}
\title{Electronic magnetism, coherence, and superconductivity in \FeTeSe}
\title{Magnetic and superconducting phases and surface state in \FeTeSe}
\title{Understanding magnetism, superconductivity, and topological surface state in \FeTeSe}
\title{Electronic magnetism and two distinct superconductivity regimes in \FeTeSe}
\title{Electronic coherence, magnetism, and superconductivity with and without topological surface state in \FeTeSe}
\title{Distinct magnetic and superconducting phases with and without topological surface state in \FeTeSe}
\title{Magnetic, superconducting, and topological surface states on \FeTeSe\ correlate with local composition}
\title{Magnetic, superconducting, and topological surface states on \FeTeSe\ }

\author{Yangmu Li}
\affiliation{Condensed Matter Physics and Materials Science Division,
Brookhaven National Laboratory, Upton, NY 11973, USA}

\author{Nader Zaki}
\affiliation{Condensed Matter Physics and Materials Science Division,
Brookhaven National Laboratory, Upton, NY 11973, USA}

\author{Vasile O. Garlea}
\affiliation{Neutron Scattering Division, Oak Ridge National Laboratory, Oak Ridge, TN 37831, USA}

\author{Andrei T. Savici}
\affiliation{Neutron Scattering Division, Oak Ridge National Laboratory, Oak Ridge, TN 37831, USA}

\author{David Fobes}
\affiliation{Condensed Matter Physics and Materials Science Division,
Brookhaven National Laboratory, Upton, NY 11973, USA}

\author{Zhijun Xu}
\affiliation{NIST Center for Neutron Research, National Institute of Standards and Technology, Gaithersburg, Maryland 20899, USA}

\author{Fernando Camino}
\affiliation{Center for Functional Nanomaterials,
Brookhaven National Laboratory, Upton, NY 11973, USA}

\author{Cedomir Petrovic}
\affiliation {Condensed Matter Physics and Materials Science Division, Brookhaven National Laboratory, Upton, NY 11973, USA}

\author{Genda Gu}
\affiliation {Condensed Matter Physics and Materials Science Division, Brookhaven National Laboratory, Upton, NY 11973, USA}

\author{Peter D. Johnson}
\affiliation{Condensed Matter Physics and Materials Science Division,
Brookhaven National Laboratory, Upton, NY 11973, USA}

\author{John M. Tranquada}
\affiliation{Condensed Matter Physics and Materials Science Division,
Brookhaven National Laboratory, Upton, NY 11973, USA}

\author{Igor A. Zaliznyak}
\email{zaliznyak@bnl.gov}
\affiliation{Condensed Matter Physics and Materials Science Division,
Brookhaven National Laboratory, Upton, NY 11973, USA}

\begin{abstract}
{\bf

The idea of employing non-Abelian statistics for error-free quantum computing ignited interest in recent reports of topological surface superconductivity and Majorana zero modes (MZMs) in FeTe$_{0.55}$Se$_{0.45}$. An associated puzzle is that the topological features and superconducting properties are not observed uniformly across the sample surface. Understanding and practical control of these electronic inhomogeneities present a prominent challenge for potential applications. Here, we combine neutron scattering, scanning angle-resolved photoemission spectroscopy (ARPES), and microprobe composition and resistivity measurements to characterize the electronic state of \FeTeSe. We establish a phase diagram in which the superconductivity is observed only at sufficiently low Fe concentration, in association with distinct antiferromagnetic correlations, while the coexisting topological surface state occurs only at sufficiently high Te concentration. We find that FeTe$_{0.55}$Se$_{0.45}$ is located very close to both phase boundaries, which explains the inhomogeneity of superconducting and topological states. Our results demonstrate the compositional control required for use of topological MZMs in practical applications.
}


\end{abstract}

\date{\today}

\maketitle
\newpage

In the nascent field of quantum computing, processors made of conventional superconducting qubits 
\cite{Castelvecci_Nature2017,Wendin_RPP2017} are already challenging classical semiconductor computers in a quest for quantum supremacy \cite{Arute_Nature2019}. The main obstacle in this effort is quantum decoherence, overcoming which requires operation in extreme isolation and at extremely low mK temperatures, and still depends on reliable error correction \cite{Arute_Nature2019}.
A proposed alternative approach involves topological quantum computing 
\cite{Nayak_RMP2008,Sarma_NPJQI2015}, which would use qubits based on topologically-protected quantum states. The idea of employing non-Abelian topological quantum states, such as Majorana zero modes (MZMs), has stimulated an intense search for such states in condensed matter systems; however, developing practical realizations has proven challenging \cite{Sarma_NPJQI2015}.

Recent studies of superconducting Fe$_{1+y}$Te$_{0.55}$Se$_{0.45}$ have reported the detection of topological surface states (TSS) by ARPES \cite{Zhang_Science2018,Rameau_2019,Zaki_2019} and MZMs within magnetic vortex cores \cite{Wang_Science2018,Zhu_Science2019,Machida_NatMat2019} and at antiphase structural domain walls \cite{Wang_Science2020} by spectroscopic-imaging scanning tunneling microscopy (SI-STM). These discoveries herald a potential breakthrough, but with substantial caveats that are open to challenge. Bound states within vortex cores are sometimes observed at finite energies \cite{Chen_NatComm2019}; even when MZMs are observed, they are not observed in all vortex cores, and their frequency of occurrence varies with magnetic field \cite{Machida_NatMat2019}. The zero-bias conductance observed for MZMs is just a fraction of the predicted quantized value ($G_0 = 2e^2/h$) and it varies with vortex position \cite{Zhu_Science2019}. Furthermore, a Josephson STM study found inhomogeneity in the superfluid density that correlated with spatial variations in the quasiparticle strength \cite{Cho_Nature2019}. 
Understanding and practical control of the observed electronic inhomogeneities present a prominent challenge, both fundamentally and technologically. Here we investigate the source of these variations and determine the conditions required for a uniform and reproducible response, aiming to reconcile the diverse observations reported so far and advance towards realization of the technological potential of topological states. 

The electronic states in \FeTeSe\ derive from hybridization between Fe $3d$ and Te $5p$ (Se $4p$) orbitals and have a complex dependence on average composition. The topological effects are a consequence of the strong spin-orbit coupling (SOC) intrinsic to the heavy Te atom \cite{Johnson_PRL2015}. The \emph{ab initio} density functional theory (DFT) predicted a topologically-nontrivial band gap for $x \approx 0.5$ but not $x=1$ \cite{Wang_PRB2015,Xu_etal_Zhang_PRL2016,Wu_etal_Hu_PRB2016}; it is this gap that leads to the TSS with a Dirac dispersion. It also enables topological superconductivity, with an MZM where a magnetic vortex core reaches the surface \cite{Xu_etal_Zhang_PRL2016}. Although DFT provided a qualitatively correct description of band inversion and topological character governed by the SOC and bond-angle-dependent Fe$3d$-Te$5p$ hybridization for some compositions \cite{Rameau_2019,Wang_PRB2015,Xu_etal_Zhang_PRL2016,Wu_etal_Hu_PRB2016,Zhang_Science2018,Zhang_NatPhys2019,Peng_etal_Ding_PRB2019}, it proved inadequate for describing the composition dependence of electronic states. The energies calculated by DFT differ from experiment by a marked shift and a re-scaling by up to a factor of 5--7, depending on the band \cite{Zhang_Science2018,Wang_PRB2015,Peng_etal_Ding_PRB2019}. This is a consequence of strong electronic correlation ($U$) and Hunds coupling effects, which are the strongest among all FeSC \cite{YinHauleKotliar_NatMat2011}. In addition to the marked band renormalization, these effects also cause an orbital-selective (OS) electron localization \cite{Zaliznyak_PRL2011,Ieki_PRB2014,Liu_etal_Shen_PRB2015,Yi_NatComm2015,Yi_NPJQM2017}, fundamentally invalidating the rigid-band approaches and favoring electronic states that compete with superconductivity \cite{Zaliznyak_PNAS2015,Zaliznyak_PRB2012,Fobes_PRL2014,Xu_etal_Xu_PRB2016}.

While DFT suggested that Fe stoichiometry, $y$, can be used to change the electronic band filling of \FeTeSe\ \cite{Xu_etal_Zhang_PRL2016} for tuning TSS, it is now recognized that this can hardly be achieved without noticeably affecting the coherence of electronic { Bloch wave states and disrupting the formation of well-defined electronic band structure \cite{Wang_PRB2015,Zhang_NatPhys2019,YinHauleKotliar_NatMat2011}. In a metal, coherent electronic bands give rise to the sharp dispersive features observed in ARPES; electron decoherence impedes such metallic character. }
In Fe-rich compositions the correlation effects are enhanced, favoring magnetism and an NSC state, which underlies the well-documented sensitivity of the superconductivity in \FeTeSe\ to Fe content \cite{McQueen_PRB2009,Liu_etal_Mao_PRB2009,Wen_etal_Tranquada_PRB2009,Bendele_PRB2010,Viennois_JSolStChem2010,Dong_PRB2011,Sun_2019}.
The OS electron decoherence/localization can be induced by small compositional changes, or temperature \cite{Zaliznyak_PRL2011,Ieki_PRB2014,Liu_etal_Shen_PRB2015,Yi_NatComm2015,Yi_NPJQM2017,Zaliznyak_PNAS2015} and is most pronounced for Fe $3d$ and in particular $3d_{xy}$ orbital derived bands, which exhibit dramatic broadening and loss of intensity in ARPES measurements 
\cite{Wang_PRB2015,Ieki_PRB2014,Liu_etal_Shen_PRB2015,Yi_NatComm2015,Yi_NPJQM2017}. These bands play a central role in superconductivity coexisting with topological band structure and TSS \cite{Zhang_Science2018,Wang_PRB2015,Xu_etal_Zhang_PRL2016,Wu_etal_Hu_PRB2016}, which therefore require that the Fe $3d$ derived bands retain coherence.


\begin{figure*}[t]
\includegraphics[width=1.\textwidth]{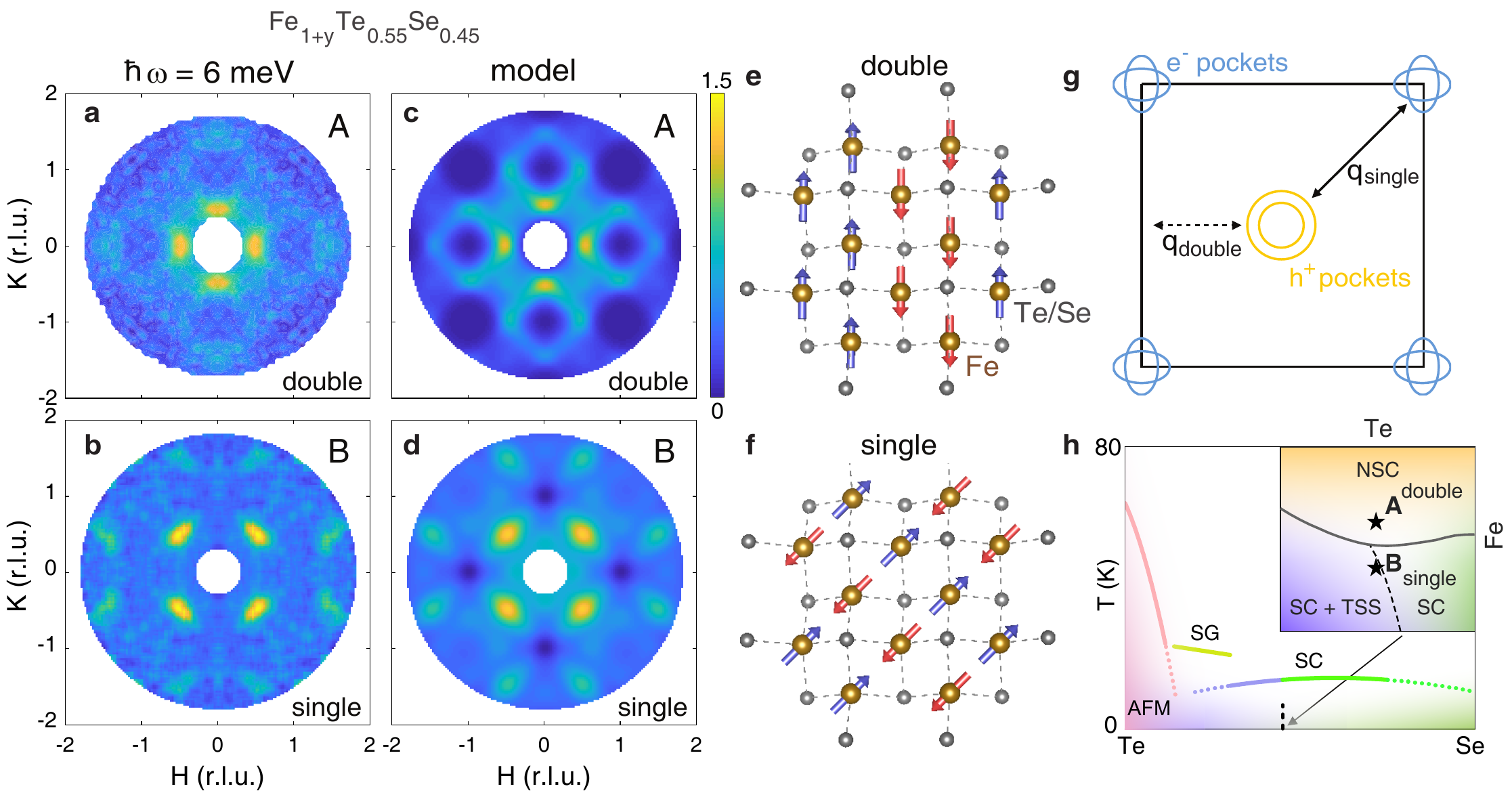}
\caption{{\bf Magnetic and superconducting phases in \FeTeSe.} (a, b) Magnetic neutron scattering measured (${\bP} || {\bQ}$ polarized, spin-flip channel) for Fe$_{1+y}$Te$_{0.55}$Se$_{0.45}$ samples A and B, respectively, at 6(1) meV energy transfer where the resonance mode \cite{Qiu_PRL2009,Thampy_PRB2014} is observed in the superconducting sample, (b). The color bar indicates absolute neutron scattering intensity from 0 to 1.5 mbarn sr$^{-1}$ meV$^{-1}$ f.u.$^{-1}$. (c, d) Model fit of the neutron scattering patterns in (a, b), respectively, using the short-range spin correlations model \cite{Zaliznyak_PNAS2015}. The scattering for the NSC sample A is similar to double-stripe magnetism observed in Fe$_{1+y}$Te$_{0.87}$S$_{0.13}$ (extended Figure~1 in \cite{Supplement}). (e, f) Illustration of the Fe-Te/Se lattice and spin arrangements for the two magnetic patterns. (g) Schematics of the Fermi surface in two-Fe unit cell indicating $q_{\rm single}$ and $q_{\rm double}$ wave-vectors corresponding to single- and double-stripe magnetism. (h) Temperature-composition and Fe-Te (insert) phase diagrams. SC, SC + TSS, NSC, AFM, and SG denote superconductivity, superconductivity with topological surface state, non-superconductive regime, antiferromagnetism, and spin glass, respectively. Stars show the average chemical composition of \FeTeSe\ samples A and B. }
\label{Fig1:neutrons_PD}
\end{figure*}

In this work, we \emph{experimentally} characterize the local compositional dependence of the electronic states in high-quality single crystals of \FeTeSe. We use polarized neutron scattering (PNS) to probe electronic magnetism and superconductivity of the bulk phase and position-resolved laser-based ARPES to measure the electronic states near the Fermi energy, $E_\mathrm{F}$, as a function of position on the cleaved crystalline surface, followed by measurements of local resistance and elemental composition at the same locations by Energy-Dispersive X-ray Spectroscopy (EDS). We find that these data can be sorted into phase diagrams as a function of local Te and Fe concentrations in which the nonsuperconducting (NSC), superconducting (SC), and SC $+$ TSS have well-defined phase boundaries and where the nonsuperconducting and superconducting phases are characterized by distinct magnetic dynamical correlations. Our central result is the experimental phase diagram of Fig.~\ref{Fig4:phase_diagram}, which is summarized in an inset of Fig.~\ref{Fig1:neutrons_PD}(h), as we explain below.

The single crystals with nominal composition Fe$_{1+y}$Te$_{0.55}$Se$_{0.45}$ and slightly different Fe content were grown by the modified Bridgman method \cite{Wen2011}. For this study, we selected two compositions (Type A and Type B), which differ only slightly in $y$ and for which large single crystals ($m \approx 20$~g) were obtained for neutron experiments. A number of small single crystals from each composition were obtained for elemental analysis and local ARPES and transport studies. While each crystal type has a well-defined average composition, $(x,y)$, sufficient spread, $(\Delta x, \Delta y)$ is observed to allow assessment of a fine-grained phase diagram. Type A samples have slightly higher Fe content ($y>0$) and show no evidence of superconductivity down to 2 K, while Type B samples are excellent bulk superconductors below $T_c = 14.5$~K \cite{Supplement}  (the difference in average Fe content for the Type A and Type B neutron samples is $\Delta y \approx 0.03$, see Fig.~S1 for more details).
The samples used in recent studies of TSS, Majorana modes, and inhomogeneous superconductivity in Fe$_{1+y}$Te$_{0.55}$Se$_{0.45}$ \cite{Zhang_Science2018,Wang_Science2018,Zhu_Science2019,Wang_Science2020,Cho_Nature2019,Zhang_NatPhys2019} are representative of our Type B crystals.

We start by considering the magnetic correlations, which were determined by polarized neutron scattering (see Methods). The magnetic responses (spin-flip scattering) for Type A and Type B crystals at an excitation energy of 6~meV and $T\lesssim 8$~K are shown in Fig.~\ref{Fig1:neutrons_PD}(a) and (b). 
{The corresponding panels, Fig.~\ref{Fig1:neutrons_PD}(c) and (d), are fits to a model of short-range correlations associated with the double-stripe and single-stripe orders shown in Fig.~\ref{Fig1:neutrons_PD}(e) and (f) \cite{Zaliznyak_PNAS2015,Supplement}. One can see that the superconducting Type B sample has antiferromagnetic correlations of the single-stripe type, consistent with other iron-based superconductors \cite{Xu_etal_Xu_PRB2016,TranquadaXuZaliznyak_JPCM2019}. In contrast, the
magnetic correlations of the nonsuperconducting Type A sample are best characterized as double-stripe type, as confirmed by analyses described in the SI (see Figs.~S2-S5).
Their magnetic energy spectra (see Fig.~S6) are markedly distinct: gapless in sample A, and with an $\approx 5$~meV gap devoid of magnetic fluctuations in superconducting sample B \cite{Supplement}.} Furthermore, the polarized neutron beam experiment confirms the magnetization results regarding superconductivity obtained on small crystals for a large sample used for neutron scattering: the Type B crystal causes full depolarization of the neutron beam on cooling below $T_c = 14.5$~K due to trapped magnetic flux associated with bulk superconductivity, whereas there is no such effect for the Type A crystal indicating no significant SC volume fraction down to 5~K \cite{Supplement,Zaliznyak_JPhys2017}.

\begin{figure*}[t!h!]
\includegraphics[width=1.\textwidth]{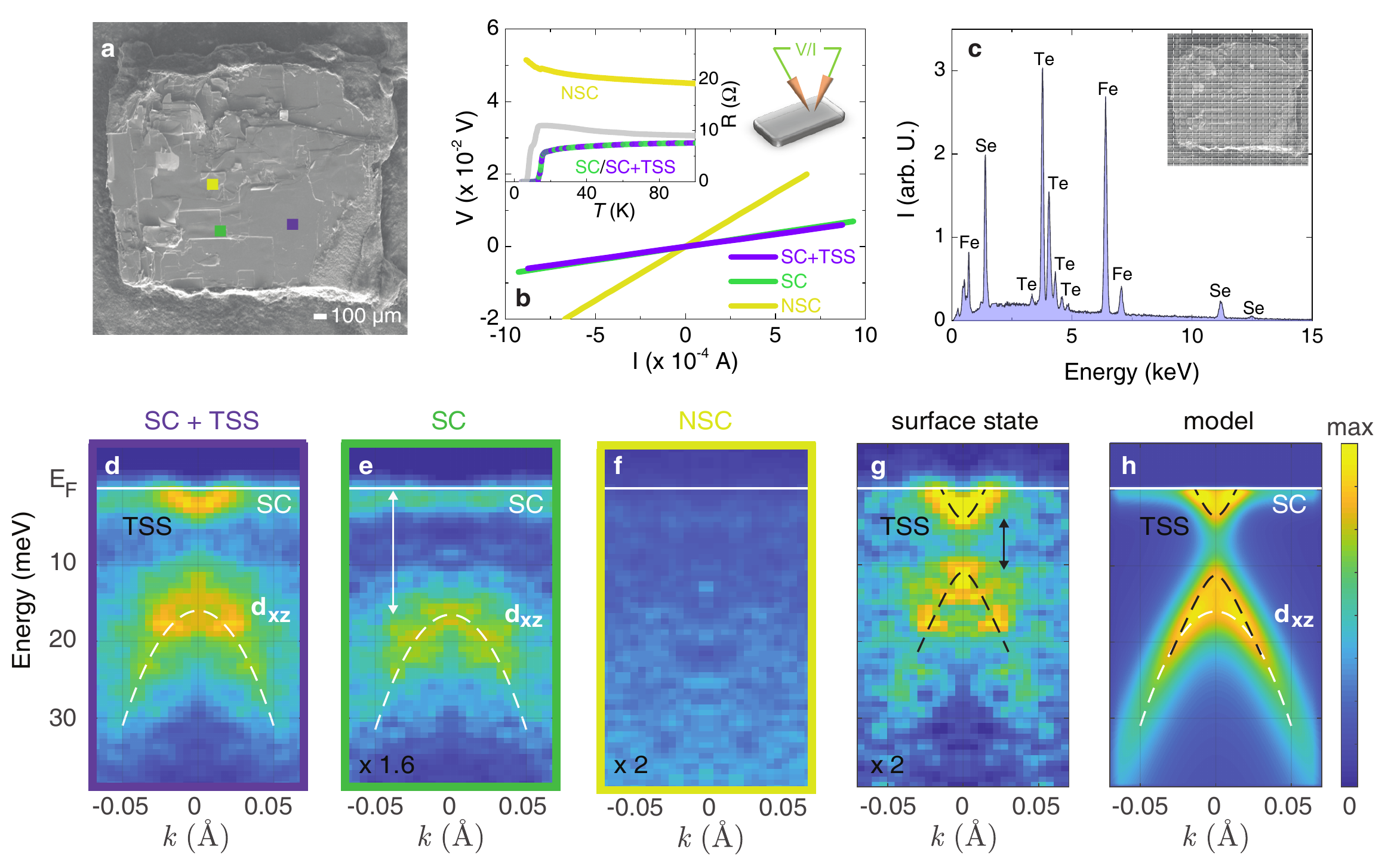}
\caption{{\bf Local electrical, chemical and photoemission properties of \FeTeSe.} (a) Scanning electron microscope image of a representative crystal piece cleaved from the sample B. (b) Representative local electrical current-voltage measurements using microprobes at 300 K. NSC, SC, and SC+TSS measurements correspond to the respective color-labeled areas in (a); the inset shows temperature dependence of the resistance measured in the respective areas. (c) Representative Energy-dispersive X-ray spectroscopy (EDS) measurement used for the compositional analysis. The inset shows the grid, which was used for all local measurements on this sample. (d, e, f) Symmetrized ARPES data at 5 K for the (d) SC+TSS, (e) SC, and (f) NSC areas color-labeled blue, green, and gold, respectively, in panel (a). White line shows the chemical potential, $E_F$. Dashed white line indicates the bulk Fe ($3d_{xz}$) band. White arrow in (e) indicates the bulk band gap, $E_F - E_B$, where $E_B$ is the energy of the top of the bulk band. (g) Photoemission intensity of TSS obtained by subtracting (e) from (d). Dashed black line shows the massive Dirac surface state; black arrow indicates the double of Dirac mass. (h) Simulated photoemission intensity using the three-component model, $I_\mathrm{B}+I_\mathrm{SC}+I_\mathrm{TSS}$, described in the text. Numbers in (f) and (g) indicate the multiplicative factor applied to photoemission intensity in order to optimize visualization. }
\label{Fig2:local_data}
\end{figure*}

The observed correlation between the bulk dynamical magnetism and superconductivity indicates that the NSC state in the Type A sample is associated with a distinct bulk electronic phase rather than with an impurity scattering by additional Fe atoms destroying the SC pairing/coherence within the same phase. The observed prominent change of local dynamical magnetism from Type A to Type B [Fig.~\ref{Fig1:neutrons_PD}(a),(c)] is fairly dramatic {(see also Figs.~S2-S6 in \cite{Supplement})}. In a common approach to itinerant magnetism, where magnetic response is attributed to nesting of Fermi-surface pockets in the Brillouin zone [Fig.~\ref{Fig1:neutrons_PD}(g)], such a transition requires an unusual switching of the Fermi surface-nesting vector from $(\pi,0)$ to $(\pi,\pi)$. However, nesting at the wave vector $(\pi,0)$ is inconsistent with calculations of the Fermi surface in \FeTeSe\ from density-functional theory \cite{Wang_PRB2015,Xu_etal_Zhang_PRL2016,Wu_etal_Hu_PRB2016}, suggesting that the double-stripe magnetism is favored by the increasing electronic correlation effects.

For the microscale characterizations, a total of 9 ($\sim 1$~mm-size) single crystals, split between Type A and Type B, were studied. Figure~\ref{Fig2:local_data} shows an image of a typical crystal, along with examples of  typical ARPES spectra, local transport measurements, and elemental analysis by EDS. For each crystal, all sets of measurements were done on the same 100~$\mathrm{\mu}$m scanning grid {[cf. inset in Fig.~\ref{Fig2:local_data}(c)]} chosen to ensure a reliable comparison of the distinct measurements. The actual spatial resolution for each technique was $\approx 20\mathrm{\mu}$m. {Hence, each technique probes an $\approx 20\times 20 \mathrm{\mu m}^2$ area approximately at the center of the same $100\times 100 \mathrm{\mu m}^2$ square pixel on the grid, but not necessarily the same exact spot. While the slight offsets introduce some noise to the correlation of the behaviors observed among different grid pixels, on the average the results are statistically well correlated, as we will see.}

Remarkably, three distinct types of local regions are clearly identified by ARPES {at 5 K}: SC with topological surface state [SC + TSS, Fig.~\ref{Fig2:local_data}(d)], SC (without TSS) [Fig.~\ref{Fig2:local_data}(e)], and regions with incoherent electronic structure [Fig.~\ref{Fig2:local_data}(f)], which we associate with the NSC phase.
For the first two cases, the photoemission spectra reveal a coherent, Fe-3$d$ derived bulk electronic band at an energy, $E_\mathrm{B}$, approximately 15~meV below $E_\mathrm{F}$ and a flat 
{ band of spectral weight from a second band that appears as a SC quasiparticle split below $E_\mathrm{F}$ by the SC gap \cite{Zhang_Science2018,Johnson_PRL2015,Zhang_NatPhys2019,Rinott_SciAdv2017}. The two bands are derived mainly from $d_{xz}$ and $d_{yz}$ states, with the bands split by hybridization with ligands and spin-orbit coupling. Only for SC + TSS regions is surface-state intensity also observed. This massive Dirac surface state, which is considered a key piece of evidence for topological superconductivity in \FeTeSe\ \cite{Zhang_Science2018,Zhang_NatPhys2019,Peng_etal_Ding_PRB2019}, is clearly revealed by directly subtracting the SC spectrum from SC + TSS spectrum (Fig.~\ref{Fig2:local_data}g).}
For quantitative analysis, we fit the total photoemission intensity (${I_\mathrm{tot}}$) to a sum of four intensity contributions, ${I_\mathrm{tot} = I_\mathrm{B}+I_\mathrm{SC}+I_\mathrm{TSS}+I_\mathrm{BG}}$, consisting of three coherent features and a constant background ($I_\mathrm{BG}$) describing the incoherent signal. The intensities of the Fe-3$d$ bulk band ($I_\mathrm{B}$) and TSS ($I_\mathrm{TSS}$) were modeled using $I = I_0 \cdot F(E) \cdot A(k, E)$, where $I_\mathrm{0}$ is a constant intensity prefactor, $F(E)$ is the Fermi function, and $A(k, E)$ is an appropriate normalized spectral function (see Methods). The intensity of the SC quasiparticle condensate (${I_\mathrm{SC}}$) peak was fit by a Gaussian function in energy. The relative intensities of the different features obtained from this analysis characterize the presence of the corresponding phases in the measured region.
{ARPES spectra for a SC+TSS region at 5K~$<T_c$ and 19K~$>T_c$ presented in Fig.~ S8 \cite{Supplement} illustrate SC and Dirac gaps opening across $T_c$ \cite{Rameau_2019,Zaki_2019}; comparison of the data with our model [Fig.~S8(d),(h)] shows that the model accurately describes all features present in ARPES spectra.}

\begin{figure*}[t]
\includegraphics[width=1.\textwidth]{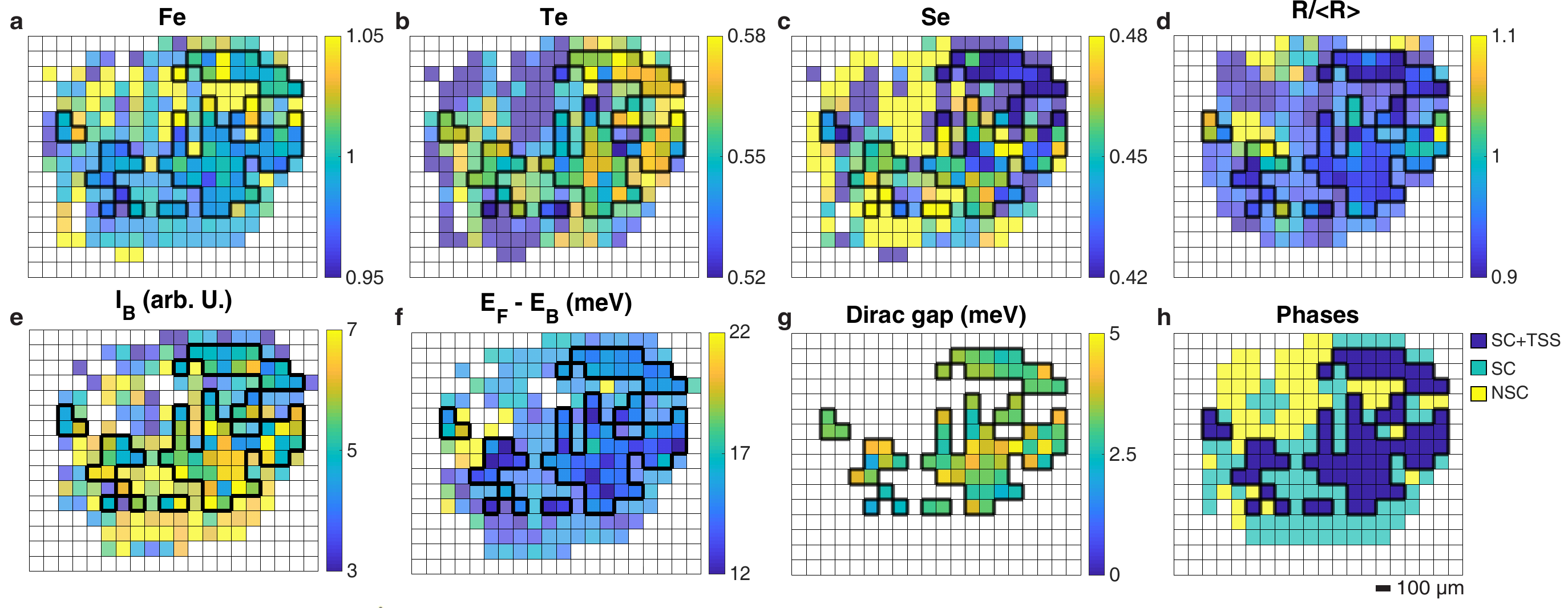}
\caption{{\bf Typical distribution of local chemical composition and electronic properties in the sample.} (a-c) Distribution of Fe, Te, and Se concentration determined from EDS measurements for the cleaved \FeTeSe\ single crystal in Fig.~\ref{Fig2:local_data}(a). Fe, Te, and Se concentrations are normalized to their nominal values. (d) Local resistance, $R$, at 300 K normalized to the average value, $\langle R\rangle$. (e-g) The integrated photoemission intensity of the bulk $d_{xz}$ band, $I_B$, bulk band gap, $E_F - E_B$, and massive Dirac gap of the topological surface state at 5 K, all obtained from ARPES measurements. (h) Spatial distribution of the NSC, SC and SC+TSS phases in the sample identified from ARPES. The measured grid is $100 \mu$m$ \times 100 \mu$m, as illustrated by the black scale bar in (h). }
\label{Fig3:phase_distribution}
\end{figure*}

The microscale voltage-current (micro-V/I) measurements provide information on the local electronic density of states and mobility. The micro-V/I measurements were performed at 300 K using 25 $\mathrm{\mu}$m-size microprobes [Fig.~\ref{Fig2:local_data}(b)]. A prominent variation of the local electrical resistance is observed for each sample, as well as between different samples. This variation, as we discuss below, is not dominated by surface irregularities or sample cleavage procedure, but is correlated with the ARPES and compositional variations. In fact, crystals of Type A (NSC), which visually have a better (shinier) surface, predominantly have higher local resistance, together with incoherent ARPES spectra. {For several representative NSC, SC, and SC+TSS regions we have also measured the temperature dependence of the resistivity, which confirm the respective NSC and SC behaviors [inset in Fig.~\ref{Fig2:local_data}(b)].

Finally, scanning-EDS measurements characterizing the local concentration of the various component elements were carried out on the same grid for all studied samples [Fig.~\ref{Fig2:local_data}(c)].}
Representative maps of the local distribution of elemental concentration and electronic properties in a typical Type B sample are shown in Fig.~\ref{Fig3:phase_distribution} (additional data for other samples are presented in Figs.~S9-S11). The maximum local variations of Fe, Te, and Se observed in our Fe$_{1+y}$Te$_{0.55}$Se$_{0.45}$ crystals are about $\pm 5\%$ of their nominal values and, as expected, there is a clear anticorrelation between Te and Se content [Fig.~\ref{Fig3:phase_distribution}(a)-(c)]. Inspection of Fig.~\ref{Fig3:phase_distribution} reveals that SC and SC + TSS regions  identified by scanning ARPES also show lower Fe content and lower local electrical resistance.
Correspondingly, the NSC regions indicated by scanning-ARPES correlate with higher resistance; the implied reduction of electronic coherence, typical of the Type A crystals, is also associated with the change of the characteristic magnetic wave vector to $(\pi,0)$ as indicated by the neutron results. Remarkably, the difference in the average Fe concentration in NSC sample A and SC sample B 
is only $\langle y\rangle_{\rm A} - \langle y\rangle_{\rm B} \lesssim 3\%$, and the full width at half maximum (FWHM) Gaussian spread of $y$ in each sample is $\lesssim 2\%$ (Fig.~\ref{Fig1:neutrons_PD}h and Fig.~S1). The superconductivity in \FeTeSe\ thus emerges in extreme proximity to {electronic} coherence-incoherence transition 
that is controlled by the small off-stoichiometry of Fe. 
We further observe that the SC + TSS regions existing within the electronically coherent SC phase exhibit a smaller energy difference between $E_\mathrm{F}$ and the top of the Fe-3$d$ bulk band ($E_\mathrm{B}$) and a higher Te concentration compared to the SC regions without TSS, corroborating the idea that the TSS is favored by stronger SOC ($\lambda$) of Te.

\begin{figure*}[t]
\includegraphics[width=1.\textwidth]{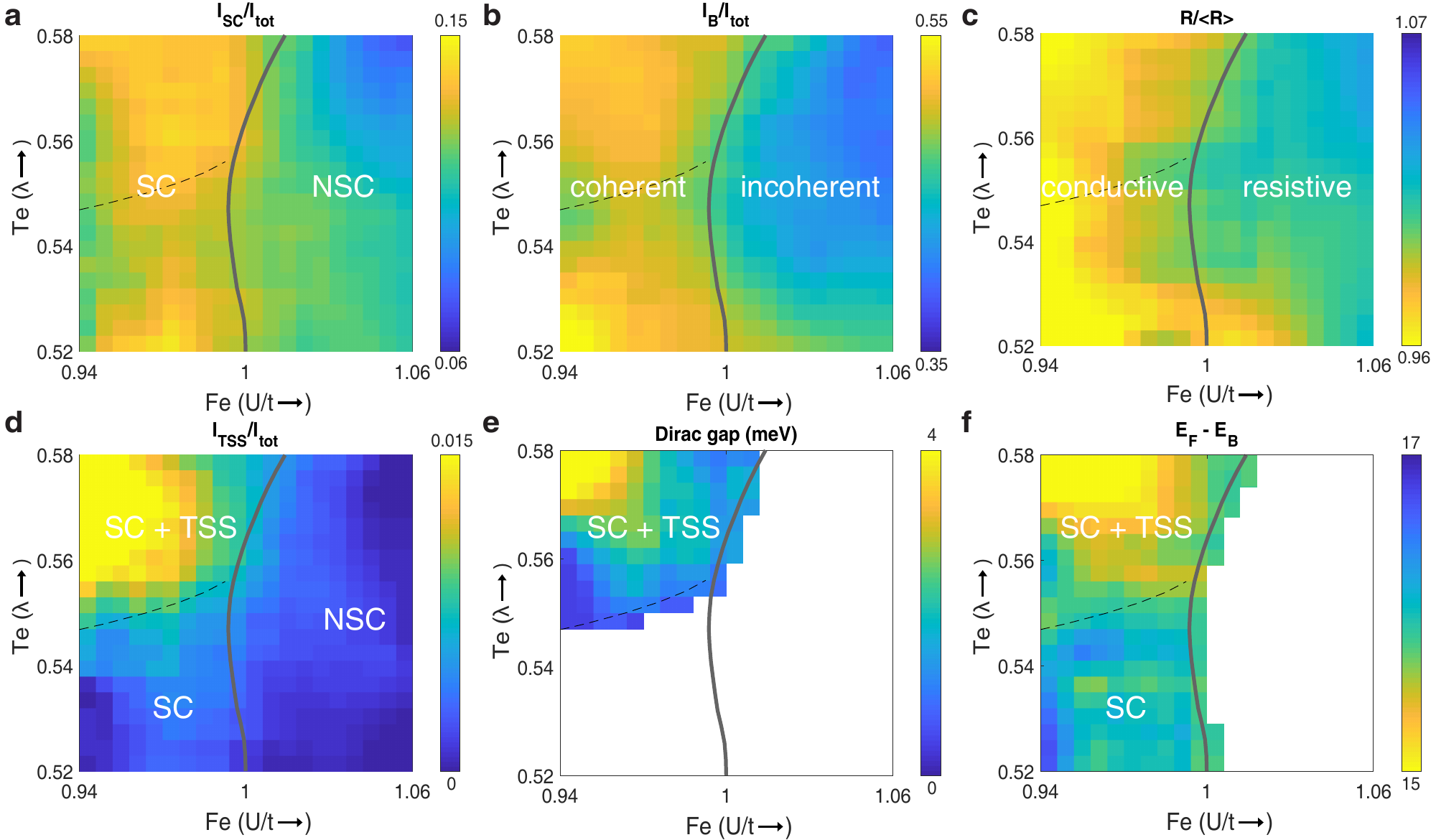}
\caption{{\bf Phase diagram of competing bulk and surface quantum states in \FeTeSe.} (a-c) Pseudocolor plot of the relative photoemission intensity at 5 K for (a) superconducting condensate, $I_\mathrm{SC}/I_\mathrm{tot}$, (b) electronic coherence quantified as the relative intensity of the coherent bulk band, $I_{B}/I_{tot}$, and (c) resistivity $R/ \langle R \rangle$ at 300 K, as a function of Fe and Te concentration. Solid curve shows the boundary between SC and NSC regime as determined based on the averaged 50-percent-threshold increase of all three quantities shown in (a-c) (also see Fig.~S12). Within our data, the onset of superconductivity is indistinguishable from the onset of electronic coherence. (d-f) Pseudocolor plots of the relative photoemission intensity for (d) topologically-protected surface state, $I_\mathrm{TSS}/I_\mathrm{tot}$, (e) Dirac gap, which can be identified in this state, and (f) the bulk band gap, $E_F - E_B$, which can be identified in the electronically coherent state, as a function of Fe and Te concentration. Dashed curve is the guide to the eye, indicating phase boundary where TSS emerges. Pseudocolor plots show data that is linearly interpolated and smoothed with Savitzky-Golay algorithm.
}
\label{Fig4:phase_diagram}
\end{figure*}

We now combine the scanning-ARPES and scanning-EDS results for 9 cleaved \FeTeSe\ crystals ($> 2000$ total grid sites; 4 pieces of Type A and 5 of Type B) with micro-V/I for 4 cleaved crystals ($\sim 1000$ grid sites; 1 piece of Type A and 3 of Type B) to create diagrams of electronic characterizations as a function of the Fe and Te concentrations, shown in Fig.~\ref{Fig4:phase_diagram}, {where phase boundaries are determined by the 50\%\ threshold of the relevant measured quantities}.
The relative photoemission intensity of superconducting condensate ($I_\mathrm{SC}/I_\mathrm{tot}$, Fig.~\ref{Fig4:phase_diagram}a), the coherent Fe-3$d$ band intensity ($I_\mathrm{B}/I_\mathrm{tot}$, Fig.~\ref{Fig4:phase_diagram}b), and local resistance ($R/ \langle R \rangle$, Fig.~\ref{Fig4:phase_diagram}c), which is sensitive to the emergence of coherent band structure needed for superconductivity, all present consistent patterns with Fe concentration, $1+y$. The experimental phase boundary between SC and NSC phases indicated by the solid curve in Fig.~\ref{Fig4:phase_diagram} represents the average for all three measured quantities (see also Fig.~S8). In addition, the TSS occurs only within the SC regime, $1+y \lesssim 1$. 
The relative photoemission intensity of  the TSS ($I_\mathrm{TSS}/I_\mathrm{tot}$, Fig.~\ref{Fig4:phase_diagram}d) exhibits a strong correlation with high Te concentration, and determines an experimental boundary of the TSS + SC phase (dashed line). The phase boundary for the TSS determined in this way is consistent with the development of the Dirac gap (Fig.~\ref{Fig4:phase_diagram}e) and with the Fe-3$d$ bulk binding energy, $E_\mathrm{F}-E_\mathrm{B}$: the disappearance of the TSS intensity with decreasing Te concentration, $1-x$, is concomitant with the closure of TSS Dirac gap and sinking of the bulk band below $\approx 16$~meV at $x \gtrsim 0.45$ (Fig.~\ref{Fig4:phase_diagram}f).

\begin{figure*}[t]
\includegraphics[width=1.\textwidth]{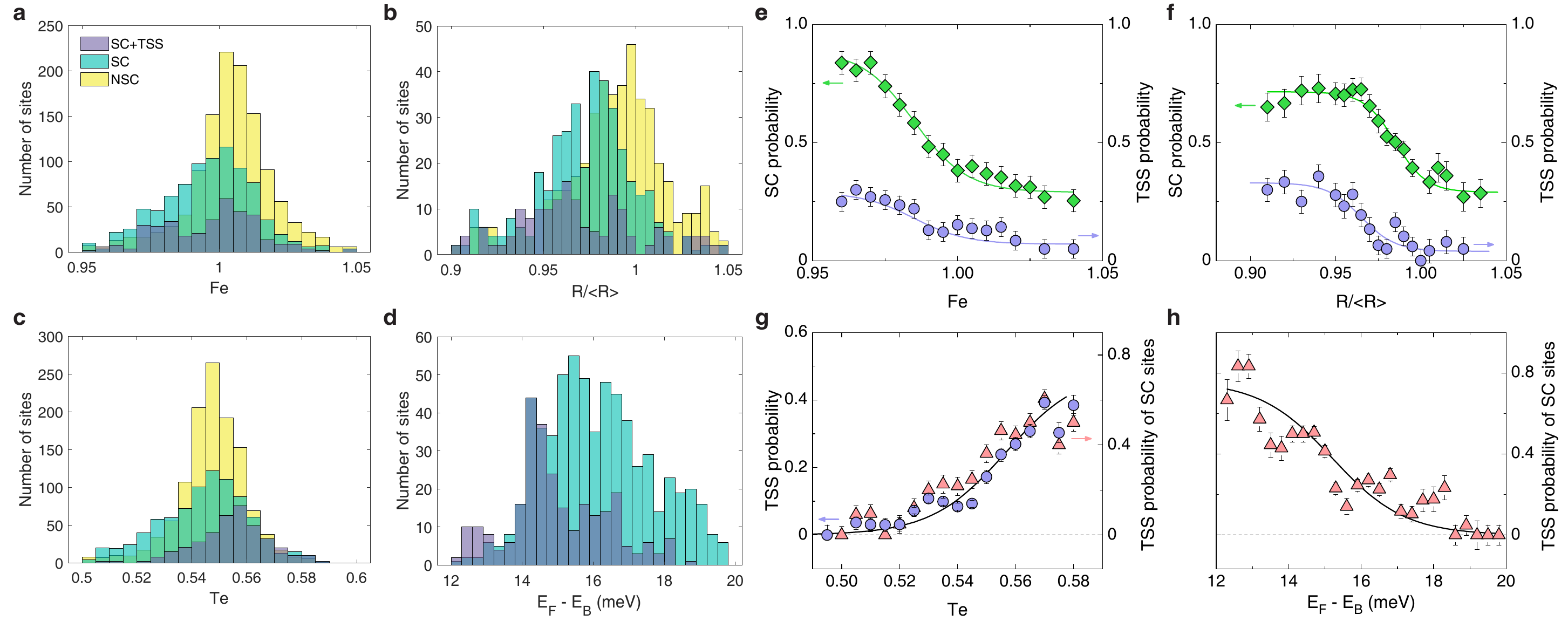}
\caption{{\bf Statistics of local phases and chemical composition in \FeTeSe\ samples.} (a-d) Histogram of the distribution of NSC, SC, and SC + TSS sites as a function of Fe concentration, $R/ \langle R \rangle$, Te concentration, and $E_F - E_B$, respectively. (e, f) SC and TSS probabilities as a function of (e) Fe concentration and (f) $R/ \langle R \rangle$. (g) TSS probability and TSS probability on SC sites as a function of Te concentration. (h) TSS probability on SC sites as a function of $E_F - E_B$. Solid lines in (e-h) are guides to the eye and dashed lines indicate zero. The statistics of local phases is obtained based on photoemission and chemical measurements for $> 2000$ sites and electrical resistance for $\approx 1000$ sites. }
\label{Fig5:phase_statistics}
\end{figure*}

The statistics of the SC + TSS, SC, and NSC local sites as a function of Fe, Te, local resistance, and $E_\mathrm{F}-E_\mathrm{B}$ are further summarized in Fig.~\ref{Fig5:phase_statistics}. The emergence of the TSS at high Te concentration within the SC phase is distinctly correlated with the Fe-3$d$ bulk band energy. As a function of Fe content, $1+y$, the emergence of the TSS is concomitant with the appearance of superconductivity (Fig.~\ref{Fig5:phase_statistics}e,f), which is observed in close proximity to (or coincident with) the electronic coherence-incoherence transition and the emergence of the coherent bulk band. Superconductivity thus appears only in the presence of electronic coherence, in a parallel to the correlation between orbital order and electron delocalization in Fe$_{1+y}$Te parent system \cite{Zaliznyak_PRL2011,Fobes_PRL2014}.

Our results show that topological superconductivity in Fe$_{1+y}$Te$_{0.55}$Se$_{0.45}$ occurs in proximity to both a coherence-driven SC-NSC phase transition and a spin-orbit-coupling-driven transition between topological (TSS + SC) and trivial SC phases. Unavoidable {inhomogeneities of the electronic order} in the critical range of these transitions (which is large for quasi-two-dimensional layered systems) \cite{ZacharZaliznyak2003,Phillabaum_NatComm2012} naturally explain the nano-scale inhomogeneity of the SC and topological properties observed here and in previous studies \cite{Zhang_Science2018,Wang_Science2018,Zhu_Science2019,Wang_Science2020,Machida_NatMat2019,Cho_Nature2019}. In agreement with experiments, this electronic inhomogeneity is controlled not by arbitrary chemical disorder, but by proximity to a phase transition line, thus reconciling and explaining the inconsistent experimental observations on Fe$_{1+y}$Te$_{0.55}$Se$_{0.45}$ \cite{Zhang_Science2018,Wang_Science2018,Zhu_Science2019,Wang_Science2020,Machida_NatMat2019}.
In our microprobe measurements, the nanoscale fluctuations are averaged out on a $\sim 10 \mu$m scale, so that a point on our phase diagram corresponds to average electronic and chemical properties.

The development of electronic incoherence, which we observe with excess Fe content, is often associated with a local Coulomb repulsion energy $U$, whereas the Te content of our \FeTeSe\ samples can be associated with spin-orbit coupling, parameterized by $\lambda$. Making such connections, it is intriguing to compare our phase diagram as a function of $(1-x,1+y)$ with the generic $(\lambda,U)$ phase diagram for quantum electronic materials discussed by Witczak-Krempa {\it et al.} \cite{Witczak-Krempa_AnnuRev2014}. The latter phase diagram does not include superconductivity, but it does contain a wider variety of topological states than have been identified so far in \FeTeSe; the comparison should inspire further investigations. From a more practical perspective, our results point to the composition range necessary to achieve well-defined properties suitable for applications in topological quantum computing. The grand challenge of realizing the technological potential of TSS on \FeTeSe\ hinges on our ability to understand and control the electronic states of this material on the level achieved in silicon technology; our work is a critical step in this direction. The next challenge is to learn to synthesize samples with the required composition in a controlled and reproducible way.

\medskip
\noindent{\bf Methods:}

{\bf{Crystal synthesis and characterization.}} High quality Fe$_{1+y}$Te$_{0.55}$Se$_{0.45}$ and Fe$_{1+y}$Te$_{0.87}$S$_{0.13}$ single crystals were synthesized with the self-flux method \cite{Wen2011}. Crystal pieces from the same batch of Fe$_{1+y}$Te$_{0.55}$Se$_{0.45}$ of Type A were previously measured by unpolarized neutron scattering \cite{Xu_PRB2010} and  X-ray scattering \cite{Liu_PRB2011}. Type B pieces have been measured by ARPES \cite{Zhang_Science2018, Ieki_PRB2014,Zhang_NatPhys2019}, STM \cite{Wang_Science2018, Wang_Science2020, Cho_Nature2019}, and optical spectroscopy techniques \cite{Homes_PRB2016}. Magnetic susceptibility measurement were performed with a MPMS XL from Quantum Design Inc.

{\bf{Neutron scattering.}} The vector-polarized (XYZ) time-of-flight neutron scattering measurements were performed at the HYSPEC spectrometer, Spallation Neutron Source. Three polarization modes ({\it i.e.}, ${\bf P} || {\bf Q}$, ${\bf P} \perp {\bf Q}$, and ${\bf P} || {\bf z}$, for momentum transfer $\hbar {\bf Q}$ and direction ${\bf z}$ perpendicular to the scattering plane) were used for Fe$_{1+y}$Te$_{0.55}$Se$_{0.45}$ ($\approx 20$~g each for both Type A and Type B, $E_i = 20$~meV).  The spin-flip cross section measured with nominal ${\bf P} || {\bf Q}$ is shown in Fig.~\ref{Fig1:neutrons_PD}; it corresponds to the dynamical spin correlations $S^{yy}+S^{zz}$ \cite{Zaliznyak_JPhys2017}.  A measurement in unpolarized mode was carried out for Fe$_{1+y}$Te$_{0.87}$S$_{0.13}$ ($\approx 2.6$~g, $E_i = 15$ meV).  This probes dynamical spin correlations $S^{xx}+S^{yy}+S^{zz}$ plus nonmagnetic background.
The  magnetic scattering patterns were fit using a 4 Fe-spin plaquette model for which the  dynamical correlation function of magnetic moments is weighted by the
plaquette structure factor, $S_p({\bf Q})=|\sum_{\nu}m_{\nu}e^{-i{\bf Q}\cdot {\bf r}_{\nu}}|^2$, where $m_{\nu}$ is the magnetic moment at the site $r_{\nu}$, and $\nu = 1, 2, 3, 4$ numbers the sites of the plaquette \cite{Zaliznyak_PNAS2015,Supplement}.

{\bf{Angle-resolved photoemission spectroscopy.}} The scanning angle-resolved photoemission spectroscopy (scanning ARPES) studies were carried out using a 3 ps pulse width, 76 MHz rep rate, Coherent Mira 900P Ti sapphire laser, the output of which was quadrupled to provide $\approx$ 6 eV incident light, and focused to a spot size of $\approx$ 20 $\mu$m in diameter. Photoemission spectra were obtained using a Scienta SES 2002 electron spectrometer. {The experimental energy resolution was 2.5 meV and the wave vector (angular) resolution was $\approx 0.002\AA^{-1}$.} The measurements used a 100 $\times$ 100 $\mu$m$^2$ grid and were obtained using $p$-polarized light in the direction perpendicular to the reflection plane. The photoemission spectra were modeled with four separate contributions - bulk band intensity ${I_\mathrm{B}}(k,E)$, the superconducting condensation intensity ${I_\mathrm{SC}}(k,E)$, the surface state intensity ${I_\mathrm{TSS}}(k,E)$, and a constant background ${I_\mathrm{BG}}$. ${I_\mathrm{B}}(k,E)$ and ${I_\mathrm{TSS}}(k,E)$ can be expressed as $I_\mathrm{0} F(E) A(k, E)$, where $I_\mathrm{0}$ is a constant, $F(E)$ is the Fermi-Dirac function, and $A(k, E)$ is the corresponding spectral function. For our experiment, we found that $A(k, E)$ can be described by
$A(k, E) = \Sigma''/\{[E-E(k)-\Sigma']^2+\Sigma''^2\}$, where $E(k)$ denotes the band dispersion, $\Sigma'=z[E-E(k)]$, and $\Sigma''= \Gamma + \gamma E^2$ with constant $z, \Gamma, \gamma$ (see Supplementary Information \cite{Supplement} for details).

{\bf{Energy-dispersive X-ray and micro voltage-current measurements.}} Scanning energy-dispersive X-ray measurements (scanning-EDS) where performed using Analytical Scanning Electron Microscope JEOL 7600F and a 80 $mm^2$ silicon drift detector with energy resolution of 129 eV for a 100 $\times$ 100 $\mu$m$^2$ scanning grid. Absolute atomic percentage of Fe, Te, Se were obtained by fitting the EDS spectra below 15 keV and then normalized to the nominal chemical concentrations using the average value over all measurement sites. Micro voltage-current measurements of cleaved crystals at $\approx$ 300~K were performed using the Signatone CM-170 microprobe station with 4 Agilent 4156C Semiconductor Parameter Analyzers. As no apparent difference is observed between data taken using 4-point and 2-point contact modes, the 2-point contact mode was used for simplicity. {The representative temperature-dependent local resistivity was measured using a similar microprobe setup with an Oxford Instruments cryostat.} The measurement used a 100 $\times$ 100 $\mu$m$^2$ grid with microprobe size of approximately 20 $\mu$m$^2$.

\begin{acknowledgments}
{\bf Acknowledgments}
We gratefully acknowledge discussions with M. Lumsden, A. Tsvelik, A. Balatsky and X. Wang, and technical assistance from F. Loeb and M. K. Graves-Brook. This work at Brookhaven National Laboratory was supported by Office of Basic Energy Sciences (BES), Division of Materials Sciences and Engineering,  U.S. Department of Energy (DOE),  under contract DE-SC0012704. Work at BNL's Center for Functional Nanomaterials (CFN) was sponsored by the Scientific User Facilities Division, Office of Basic Energy Sciences, U.S. Department of Energy, under the same contract. This research used resources at the Spallation Neutron Source, a DOE Office of Science User Facility operated by the Oak Ridge National Laboratory.  \\
%
{\bf{Author contributions: }}
I.Z. conceived and directed the study. I.Z. and Y.L. designed the study with the input from J.T. and P.J.; G.G., Z.X., and C.P. provided the samples for the study. I.Z., Y.L., D.F., A.S. and O.G. carried out neutron experiments and obtained the neutron data. N.Z. carried out ARPES measurements and initial ARPES analysis; Y.L. performed fitting of ARPES spectra. Y.L. performed local EDS and microprobe measurements with help from F.C. Y.L. and I.Z. analyzed the data, prepared the figures and wrote the paper, with input from all authors.
%
{\bf{Competing Interests: }}
The authors declare that they have no competing interests.
%
{\bf{Data availability: }}
All data needed to evaluate the conclusions in the paper are present in the paper and/or the Supplementary Materials. Additional data available from authors upon reasonable request.

\end{acknowledgments}



\pagebreak
\hypersetup{pageanchor=false}
\renewcommand{\thepage}{S\arabic{page}}
\setcounter{page}{1}
\renewcommand{\theequation}{S\arabic{equation}}
\setcounter{equation}{0}
\renewcommand{\thefigure}{S\arabic{figure}}
\setcounter{figure}{0}

\section*{Supplementary Information}
\begin{center}
{\bf Magnetic, superconducting, and topological surface states on \FeTeSe} \\
Y.~Li, N.~Zaki, A.~T.~Savici, O.~V.~Garlea, D.~Fobes, Z.~Xu, F.~Camino, C.~Petrovic, G.~D.~Gu, P.~D.~Johnson, J.~M.~Tranquada, and I.~A.~Zaliznyak\\
correspondence to: zaliznyak@bnl.gov
\end{center}

\subsection{Crystal synthesis and characterization}
\label{CrystalSynthesis}
High quality Fe$_{1+y}$Te$_{0.55}$Se$_{0.45}$ and Fe$_{1+y}$Te$_{0.87}$S$_{0.13}$ single crystals were synthesized with the self-flux method \cite{Wen2011}. Crystal pieces from the same batch of Fe$_{1+y}$Te$_{0.55}$Se$_{0.45}$ of Type A were previously measured by unpolarized neutron scattering \cite{Xu_PRB2010} and X-ray scattering \cite{Liu_PRB2011}. Type B pieces have been measured by ARPES \cite{Zhang_Science2018, Ieki_PRB2014,Zhang_NatPhys2019}, STM \cite{Wang_Science2018, Wang_Science2020, Cho_Nature2019}, and optical spectroscopy techniques \cite{Homes_PRB2016}.

In the present work, two large ($\approx 20$~g) Fe$_{1+y}$Te$_{0.55}$Se$_{0.45}$ single crystals, one of Type A and one of Type B, and a Fe$_{1+y}$Te$_{0.87}$S$_{0.13}$ ($\approx 2.6$~g) single crystal, which was previously measured on a triple axis neutron spectrometer \cite{Zaliznyak_PNAS2015}, were studied by neutron scattering at HYSPEC. Nine small crystals, samples 1 through 5 of Type B (SC) and samples 6 through 9 of Type A (NSC), were used for ARPES, EDS, resistivity, and magnetic susceptibility measurements. Magnetic susceptibility measurements were performed with a MPMS XL from Quantum Design Inc.
The temperature dependence of magnetic susceptibility on a Type A (NSC) and a Type B (SC) sample confirming the corresponding NSC and SC behavior are shown in Fig.~\ref{FigS1} (d) and (h), respectively.
%

\subsection{Scanning energy-dispersive X-ray and micro voltage-current measurements}
\label{EDSandVI}
Scanning energy-dispersive X-ray measurements (scanning-EDS) where performed using Analytical Scanning Electron Microscope JEOL 7600F and a 80 $mm^2$ silicon drift detector with energy resolution of 129 eV for a 100 $\times$ 100 $\mu$m$^2$ scanning grid. Absolute atomic percentage of Fe, Te, and Se were obtained by fitting the EDS spectra below 15 keV and then uniformly cross-normalized to obtain the nominal average chemical composition, Fe$_{0.985}$Te$_{0.55}$Se$_{0.45}$, of Type B sample when using the average values over all measured sites on Type B crystals.

\begin{figure}[p!]
\includegraphics[width=1.\textwidth]{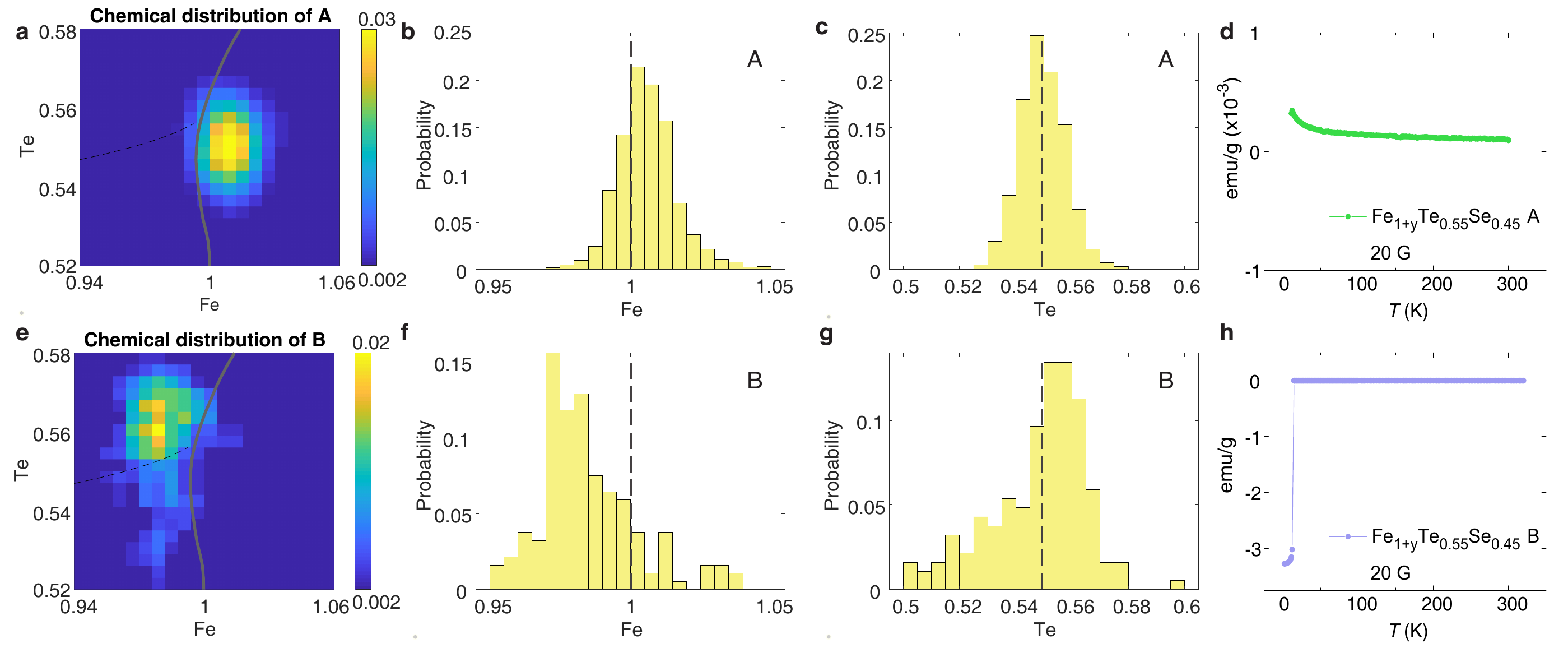}
\caption{{\bf The chemical distribution and magnetization of \FeTeSe\ Type A and Type B neutron samples. } (a-c) Pseudocolor plot and histogram of Fe and Te concentration and (d) temperature dependence of magnetization for \FeTeSe\ neutron sample A. (e-g) Pseudocolor plot and histogram of Fe and Te concentration and (h) temperature dependence of magnetization for \FeTeSe\ neutron sample B. Vertical dashed lines in (b, c, f, g) indicate the approximate phase boundaries in (a, e).}
\label{FigS1}
\end{figure}

The probability distribution of Fe and Te concentration in Type A and Type B samples obtained by combining the results from all measured crystals are shown in the form of pseudocolor maps in panels (a) and (e), respectively, of Fig.~\ref{FigS1}. The corresponding probability histograms are shown in panels (b), (c) [Type A] and (f), (g) [Type B] of Fig.~\ref{FigS1}. These results place the bulk of Type A and Type B crystals on the two sides of the NSC-SC phase transition line (solid line) determined from the combination of ARPES and resistivity measurements as described in the main text. The results also place bulk of the Type B sample in SC+TSS phase (broken line), but with a sizeable fraction of sites on the tail of the probability distribution that are within a topologically trivial SC phase at lower Te concentration.

Micro voltage-current measurements of cleaved crystals were performed using the Signatone CM-170 microprobe station with 4 Agilent 4156C Semiconductor Parameter Analyzers at $\approx$ 300 K. As no apparent difference is observed between data taken using 4-point and 2-point contact modes, the 2-point contact mode was used for simplicity. {The representative temperature-dependent local resistivity was measured using a similar microprobe setup with an Oxford Instruments cryostat.} The measurement used a 100 $\times$ 100 $\mu$m$^2$ grid with microprobe size of approximately 20 $\mu$m$^2$. The resistivity results for additional three small crystals, 1 (Type B), 2 (Type B) and 9 (Type A) are presented in Figs.~\ref{FigS5}(d), \ref{FigS6}(d), and \ref{FigS7}(d), respectively. The phase diagram obtained from the resistivity measurements is presented in Fig.~\ref{FigS8}(c). The black dashed line shows the experimental phase boundary corresponding to a 50\% change of the resistivity as a function of $y$; white dashed lines overlaid on the resistivity color map show phase boundaries obtained from ARPES measurements as described in the main text.

\subsection{Polarized neutron scattering measurements}
\label{NeutronScattering}
The vector-polarized (XYZ) time-of-flight neutron scattering measurements were performed at the HYSPEC spectrometer, Spallation Neutron Source, Oak Ridge National Laboratory. Three polarizations of the incident neutron beam ($E_i = 20$~meV) were used for measurements on both Type A and Type B Fe$_{1+y}$Te$_{0.55}$Se$_{0.45}$ samples: (i) ${\bP} || {\bQ}$, with neutron spin in the scattering plane and parallel to the wave vector transfer, ${\bQ}$, for elastic scattering at the center of the detector bank \cite{Zaliznyak_JPhys2017}, (ii) ${\bP} \perp {\bQ}$, with neutron spin in the scattering plane and perpendicular to the wave vector transfer ${\bQ}$, and (iii) ${\bP} || {\hat{z}}$, with neutron spin in the vertical direction, perpendicular to the scattering plane. An additional unpolarized neutron measurements ($E_i = 15$~meV) were carried out at HYSPEC for both Fe$_{1+y}$Te$_{0.55}$Se$_{0.45}$ and Fe$_{1+y}$Te$_{0.87}$S$_{0.13}$ samples. In all cases crystals were attached to an aluminum sample holder and mounted on a cold flange in closed-cycle refrigerator; intensity from non-sample-related scattering was minimized using neutron absorbers.

{We have also investigated an NSC Type C sample with higher Fe content than Type A sample, but PNS indicated the presence of Fe inclusions which depolarized the beam, suggesting multiple chemical phases. We therefore measured Type A and Type B samples where no such effects were observed. 
Sensitivity of PNS to depolarizing field of ferromagnetic inclusions could be used to put an upper limit of $\lesssim 0.1\%$ on possible volume fraction of Fe inclusions in the sample (the volume magnetization equal to $\sim 10^{-3}$ of Fe remanent magnetization, $\approx 0.6$~T, is of the order of neutron guide field on the instrument and would cause the depolarization of neutron beam). This allows to corroborate macroscopic chemical Fe stoichiometry of our large neutron samples.

Similarly, a rough estimate of the superconducting fraction can be obtained for a SC sample from beam depolarization. Assuming a homogeneous distribution of SC and NSC fractions in the sample that neutron beam passes through and associating the probability for a neutron to depolarize with the probability of finding a SC volume in the sample, a 50\% SC volume fraction would correspond to full depolarization of 50\% of all neutrons and would reduce the flipping ratio (FR) from $\approx 14$ to $\approx 3$. Upon cooling sample B in horizontal guide field, we observed full beam depolarization with FR~$\lesssim 1.2$, for which the above estimate suggests SC volume fraction $\gtrsim 90\%$. Cooling the sample in zero magnetic field environment reduces frozen magnetic field in the SC state. For sample B, this procedure allowed to obtain FR varying, as a function of sample rotation, in the 3 to 8 range \cite{Zaliznyak_JPhys2017}. Thanks to the fact that SF and NSF elastic Bragg scattering from crystal lattice, which contains information about FR, is contained in the measured data, the data can be straightforwardly corrected for the finite FR.}

The differential cross-section of magnetic neutron scattering can be written as,
\begin{align*}
\frac{d^{2}\sigma(\bQ,E)}{d\Omega dE} = \left(\frac{\gamma r_0}{2\mu_B}\right)^2 |\langle m_f|\bsigma \cdot \bM_{\perp}({\bQ})|m_i\rangle|^2\delta(E-\Delta E),
\end{align*}
where $\gamma = g_{n}/2$ ($g_{n} = -3.826$ is neutron g-factor), $r_0$ is the classical electron radius, $\mu_B$ is Bohr magneton, $m_f$ and $m_i$ are the final and the initial neutron spin states, respectively, and $\bsigma$ is neutron spin Pauli operator. $\hbar\bQ$ is the neutron momentum transfer and $\Delta E$ is the neutron energy transfer. The projection of the electronic magnetic moment operator given by the double cross product, $\bM_{\perp} (\bQ) = \hat{Q} \times\left[\bM({\bQ})\times \hat{Q}\right]$, $\hat{Q} = \bQ/Q$, is a property of the dipole-dipole interaction of neutron with magnetic moments in the material and is such that neutron scattering cross-section only measures moments that are perpendicular to the neutron wave vector transfer, $\bQ$; the matrix element ensures that $\bM_{\perp} (\bQ)$ is also perpendicular to $m_i$. In our polarized neutron scattering experiments, for neutron spin-flip (SF) and non-spin-flip (NSF) channels with ${\bP} || {\bQ}$, ${\bP} \perp {\bQ}$, and ${\bP} || {\hat{z}}$, the measurements can be summarized as follows,
\vspace{2mm} \\
\begin{tabular}{lll}
${\bP} || {\bQ}$: & {SF}: $S^{xx} +S^{yy}$ & {NSF}: {background} \\
${\bP} \perp {\bQ}$: & {SF}: $S^{zz}$ & {NSF}: $S^{yy}$+ {background} \\
${\bP} || {\hat{z}}$: & {SF}: $S^{xx}$ & {NSF}: $S^{zz}$+ {background} \\
\end{tabular}
\vspace{2mm} \\
where $S^{xx}$, $S^{yy}$, and $S^{zz}$ denote two-point correlation of Fe magnetic moments along $\bQ$ and perpendicular to $\bQ$ in the scattering plane and perpendicular to the scattering plane, respectively.

Representative polarized time-of-flight neutron spectroscopy data is shown in Fig.~\ref{FigS2}. Note, that because each spectrum provides information for a wide $(\bQ, E)$ coverage, $S^{xx}$, $S^{yy}$, and $S^{zz}$ should be associated with specific $(\bQ, E)$ positions, while the polarization channels, ${\bP} || {\bQ}$ and ${\bP} \perp {\bQ}$, in our notation refer to the wave vector transfer, ${\bQ}$, for elastic scattering at the center of the detector bank \cite{Zaliznyak_JPhys2017}. This explains the appearance of magnetic signal in the NSF ${\bP} || {\bQ}$ measurement of Fig.~\ref{FigS2}(h), which is in addition to $(2,0,0)/(0,2,0)$ acoustic phonon scattering at large $\bQ$. Phonon scattering is absent in SF channel.

The  magnetic scattering patterns were fit using a 4 Fe-spin plaquette model (Fig.~\ref{FigS3}), for which the dynamical correlation function of magnetic moments is weighted by the plaquette structure factor \cite{Zaliznyak_PNAS2015},
\begin{align*}
 S_p(\bQ) = \left| \sum_{\nu} \mu_{\nu}e^{-i \bQ\cdot \br_{\nu}} \right|^2 ,
\end{align*}
where $\bQ$ is the wave vector and $\mu_{\nu}$ is magnetic moment ($\pm \mu$) at the position $\br_{\nu}$ in a plaquette unit cell; index $\nu =1, 2, 3, 4$ numbers the
sites of the plaquette. For the singe- and double-stripe magnetic order, the structure factor averaged with respect to $C_4$ symmetry of plaquette orientation on the square lattice is given by,
\begin{align*}
 S_p(\bQ)=4\mu^2(\sin^2\pi h + \sin^2\pi k)(\sin^2\pi \frac{h+k}{2}+\sin^2\pi \frac{h-k}{2}) ,
\end{align*}
with propagation wave vector $\bQ_\mathrm{single}=(0.5,0.5)$ and $\bQ_\mathrm{double}=(0.5,0)$, respectively.

For the short-range inter-plaquette correlations, a set of Lorentzian peaks are expected periodically in the reciprocal lattice of the non-Bravais square lattice with 4 Fe per unit cell, which replace Bragg peaks existing in the case of magnetic long-range order. Using the factorized Lorentzian model for the peak broadening, the total structure factor can be written as,
\begin{align*}
 S(\bQ)=S_p(\bQ) \frac{\sinh 2\xi^{-1}}{\cosh\xi^{-1}+\cos(2\pi(h+k))}\frac{\sinh 2\xi^{-1}}{\cosh\xi^{-1}+\cos(2\pi(h-k))}\frac{\sinh 2\xi_c^{-1}}{\cosh\xi_c^{-1}+\cos(2\pi l)} ,
\end{align*}
where $\xi$ is the inter-plaquette correlation length in the Fe $ab$-plane and $\xi_c$ is the inter-plane correlation length \cite{Zaliznyak_PNAS2015}.

\subsection{Scanning angle-resolved photoemission spectroscopy}
\label{ARPES}
The ARPES studies reported here were carried out using a 3 ps pulse width, 76 MHz rep rate, Coherent Mira 900P Ti sapphire laser, the output of which was quadrupled to provide $\sim$ 6 eV incident light, and focused to a spot size of $\sim$ 20 $\mu$m in diameter. Photoemission spectra were obtained using a Scienta SES 2002 electron spectrometer. {The experimental energy resolution was 2.5 meV and the wave vector (angular) resolution was $\approx 0.002\AA^{-1}$.} The measurements used a 100 $\times$ 100 $\mu$m$^2$ grid and were obtained using p-polarized light in the direction perpendicular to the reflection plane. Three distinctive spectra were obtained for measurements over 3,000 sites.

The photoemission spectra (Fig.~\ref{FigS4},~\ref{FigS4.1}) were analyzed using a phenomenological model, for which the total photoemission intensity, ${I_\mathrm{tot}(k,E)}$, was separated into the bulk band intensity, ${I_\mathrm{B}(k,E)}$, the superconducting condensate intensity, ${I_\mathrm{SC}(k,E)}$, the surface state intensity, ${I_\mathrm{TSS}(k,E)}$, and a constant background, ${I_\mathrm{BG}}$. The photoemission spectral function for band electrons is described by a standard expression,
\begin{align*}
A(k, E)=\Sigma''/\pi\{[E-E(k)-\Sigma']^2+\Sigma''^2\} ,
\end{align*}
where $E(k)$ denotes the band dispersion, $\Sigma'=z[E-E(k)]$, and $\Sigma''= \Gamma + \gamma E^2$ with constant $z$, $\Gamma$, and $\gamma$.

The photoemission intensity of the bulk $d_{xz}$ band is given by,
\begin{align*}
&I_\mathrm{B}(k,E)=I_\mathrm{B}\cdot F(E) \cdot A_B(k, E)= I_\mathrm{B}\cdot F(E) \cdot \Sigma_B''/\pi\{[E-E_B(k)-\Sigma_B']^2+\Sigma_B''^2\} \\
&E_\mathrm{B}(k)=E_\mathrm{B}(k=0) + v^2k^2 ,
\end{align*}
where $E_\mathrm{B}(k)$ is the bulk band dispersion, $E_\mathrm{B}(k=0)$ is the energy at the band center, $v$ is the band velocity, $I_\mathrm{B}$ is a constant intensity prefactor, and $F(E)$ is the Fermi function.

Similarly, the photoemission intensity of the surface state intensity is given by,
\begin{align*}
&I_\mathrm{TSS}(k,E)=I_\mathrm{TSS}\cdot F(E) \cdot A_{TSS}(k, E)= I_\mathrm{TSS}\cdot F(E) \cdot \Sigma''/\pi\{[E-E_\mathrm{TSS}(k)-\Sigma']^2+\Sigma''^2\} \\
&E_\mathrm{TSS}(k)=E_\mathrm{TSS}(k=0)+\sqrt{\lambda^2k^2+E_\mathrm{Dirac}^2} ,
\end{align*}
where $E_\mathrm{TSS}(k)$ is the surface band dispersion, $E_\mathrm{TSS}(k=0)$ is the energy at the band center, $\lambda$ is the Dirac band velocity parameter, and $E_\mathrm{Dirac}$ is the gap of Dirac dispersion. $I_\mathrm{TSS}$ is a constant intensity prefactor.

The photoemission intensity of the superconducting condensate is modeled by a Gaussian peak,
\begin{align*}
&I_\mathrm{SC}(k,E)=I_\mathrm{SC}\cdot F(E) \cdot e^{-E^2/2\sigma^2} ,
\end{align*}
where $I_\mathrm{SC}$ is a constant intensity prefactor, and $\sigma\sqrt{8\log2}$ is the full-width at half-maximum (FWHM) of the intensity distribution.

In our fits, we obtained $v \approx  70 {\AA}^{-1}$, $\Gamma/(1-z) \approx 2.5$~meV, $\gamma/(1-z)$ in the range of 0.005/meV to 0.008/meV, $E_\mathrm{TSS}(k=0) = 8(1)$~meV, and $\sigma\sqrt{8\log2} \approx 4$~meV.

\begin{figure}[p!]
\includegraphics[width=1.\textwidth]{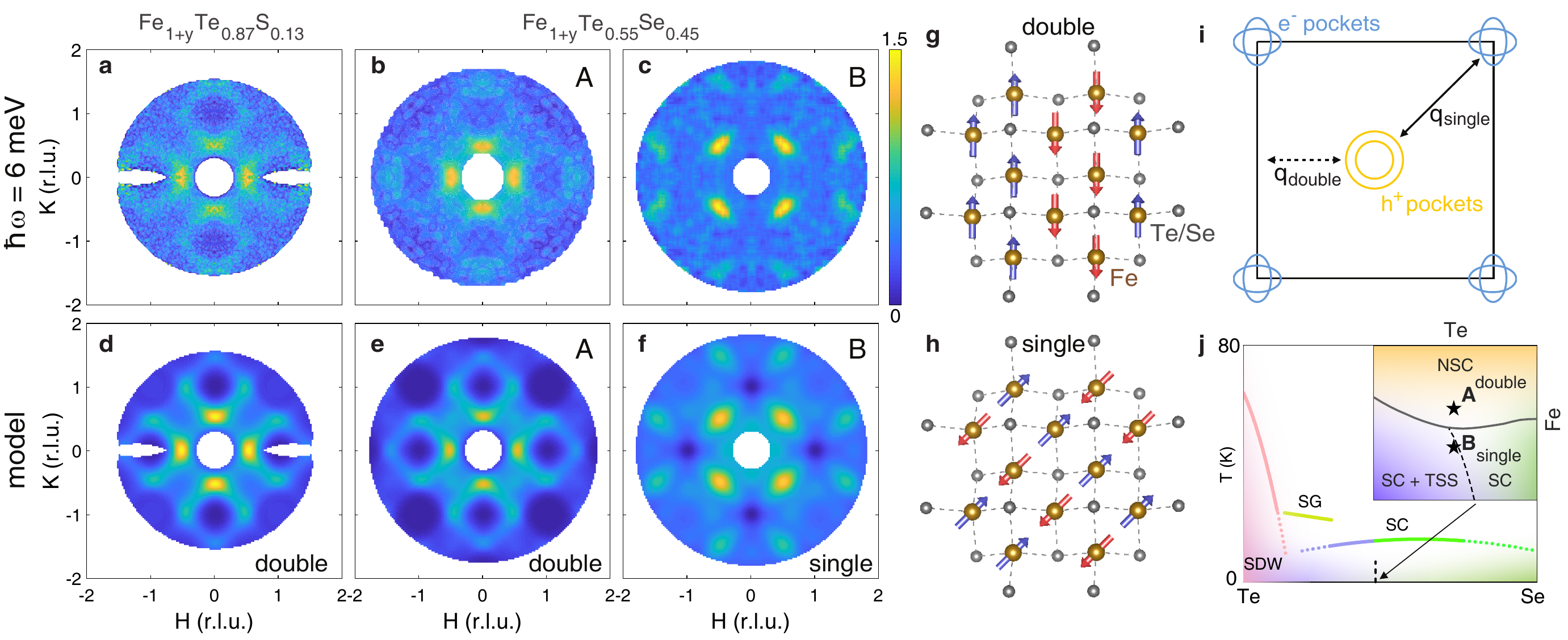}
\caption{{\bf Magnetic and superconducting phases in \FeTeSe\ and \FeTeS\ [extended Figure 1].} (a-c) Magnetic neutron scattering patterns for Fe$_{1+y}$Te$_{0.87}$S$_{0.13}$ (unpolarized) and Fe$_{1+y}$Te$_{0.55}$Se$_{0.45}$ (${\bP} || {\bQ}$ polarized, spin-flip channel) samples at 6(1) meV energy transfer where the resonance mode \cite{Qiu_PRL2009,Thampy_PRB2014} is observed in the superconducting sample, (c). The color bar indicates absolute neutron scattering intensity from 0 to 1.5 mbarn sr$^{-1}$ meV$^{-1}$ f.u.$^{-1}$. (d-f) Model fit of the neutron scattering patterns in (a-c), respectively, using the short-range spin correlations model \cite{Zaliznyak_PNAS2015}. (g, h) Illustration of the Fe-Te/Se lattice and spin arrangements for the two magnetic patterns. (i) Schematics of the Fermi surface in two-Fe unit cell indicating $q_{\rm single}$ and $q_{\rm double}$ wave-vectors corresponding to single- and double-stripe magnetism. (j) Temperature-composition and Fe-Te (insert) phase diagrams. SC, SC + TSS, NSC, AFM, and SG denote superconductivity, superconductivity with topological surface state, non-superconductive regime, antiferromagnetism, and spin glass, respectively. Stars show the average chemical composition of \FeTeSe\ samples A and B. }
\label{FigS2.1}
\end{figure}
\begin{figure}[p!]
\includegraphics[width=.8\textwidth]{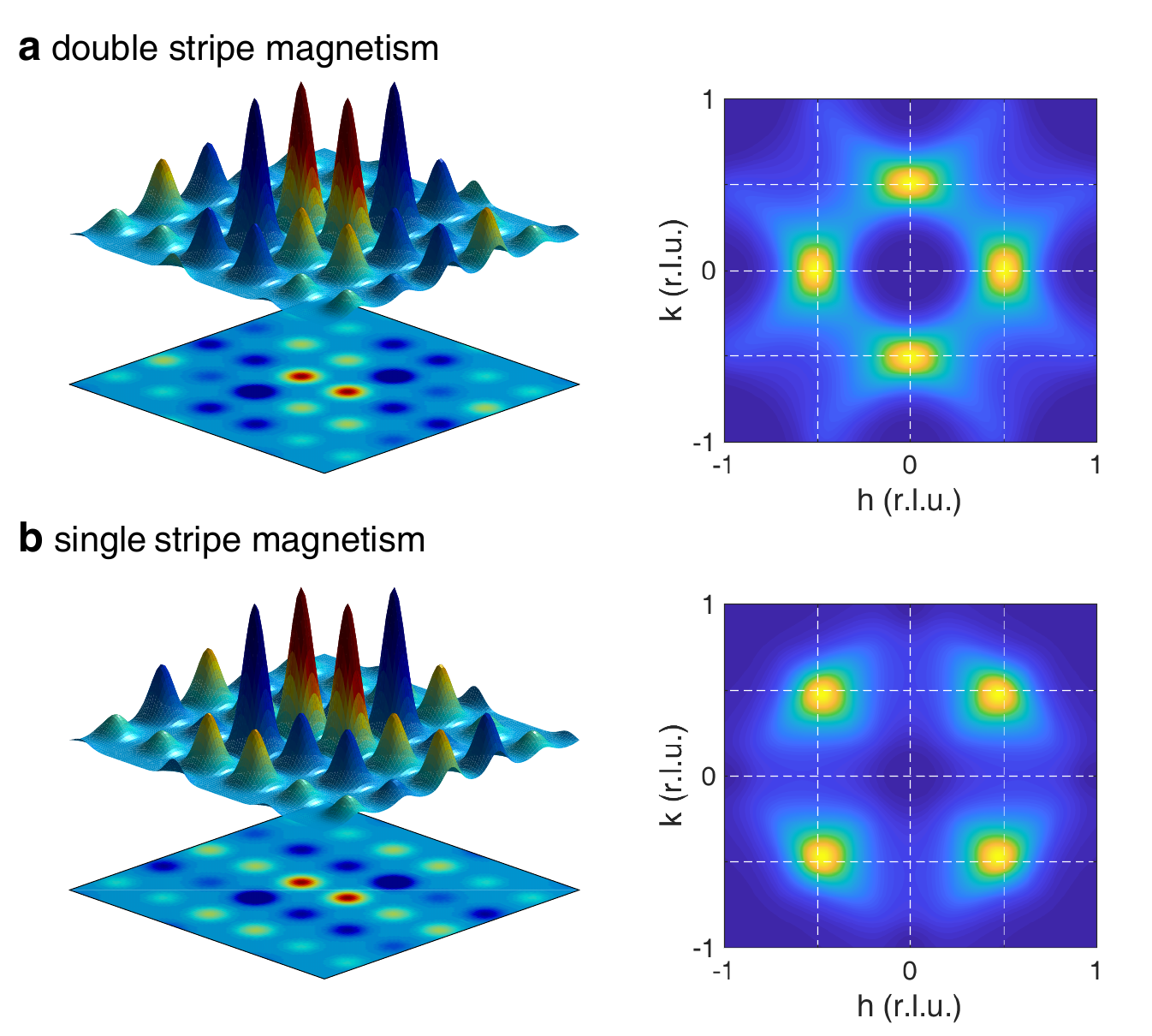}
\caption{{\bf Simulation of the neutron scattering patterns with the 4 Fe-spin plaquette model.} (a, b) Color-coded spin-spin correlation (left panels) and corresponding neutron scattering patterns (right panels) for double and single stripe magnetism. Red and blue in the left panels indicate magnetic spin moments along up and down directions.}
\label{FigS3}
\end{figure}
\begin{figure}[p!]
\includegraphics[width=.8\textwidth]{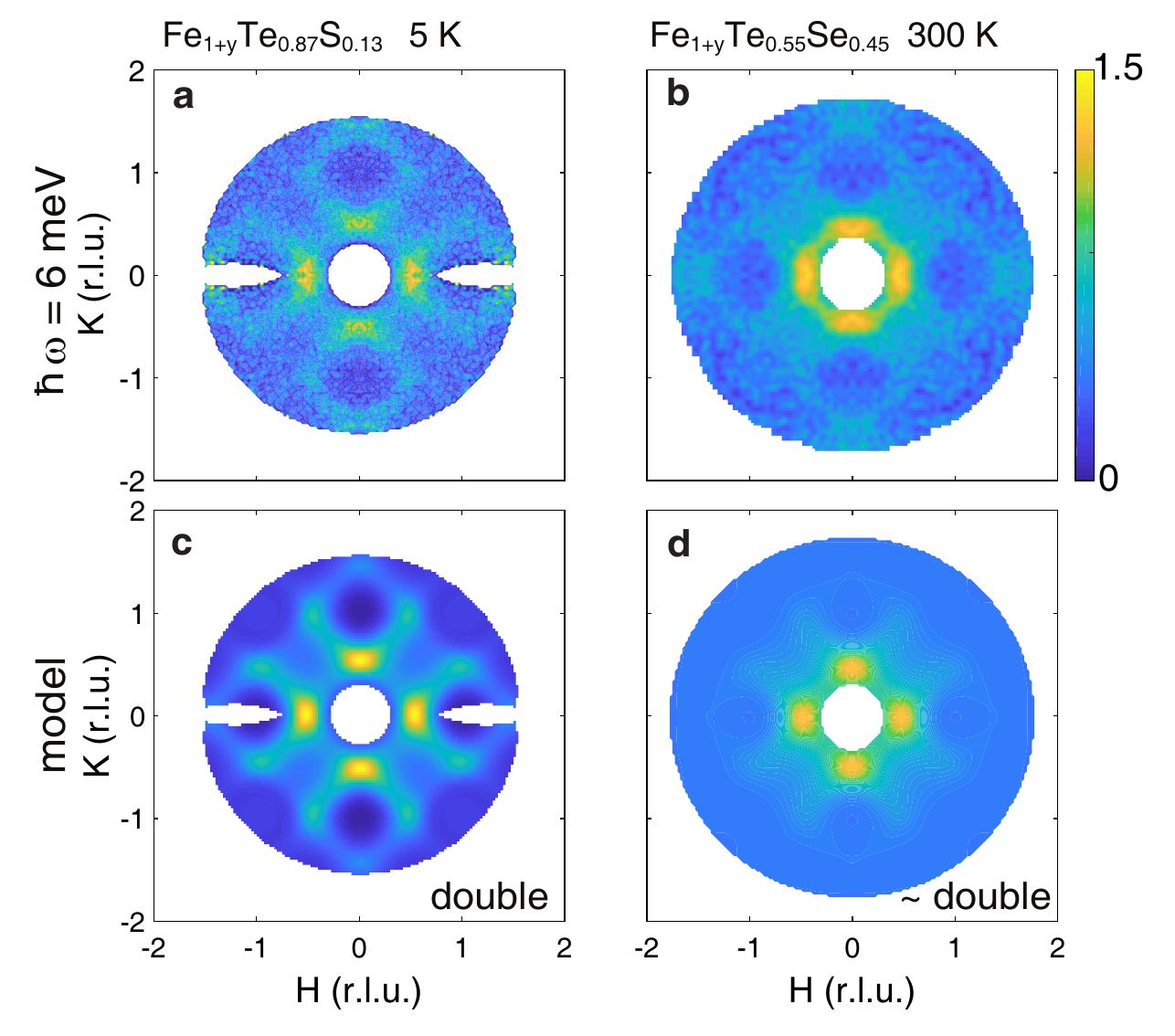}
\caption{{\bf Double-stripe magnetic scattering pattern in \FeTeS\ and NSC \FeTeSe.} (a) Magnetic neutron scattering (unpolarized) measured for Fe$_{1+y}$Te$_{0.87}$S$_{0.13}$ sample at 6(1) meV energy transfer at T = 5 K, same as in Fig.~\ref{FigS2.1}(a). (b) Magnetic neutron scattering (${\bP} || {\bQ}$ polarized, spin-flip channel) measured for Fe$_{1+y}$Te$_{0.55}$Se$_{0.45}$ sample A at the same energy transfer, 6(1) meV, at T = 300 K. (c, d) Model fit of the neutron scattering patterns in (a, b), respectively, using the short-range spin correlations model \cite{Zaliznyak_PNAS2015}. Note that patterns are similar and consistent with double-stripe magnetism.}
\label{FigS2.2}
\end{figure}
\begin{figure}[p!]
\includegraphics[width=1.\textwidth]{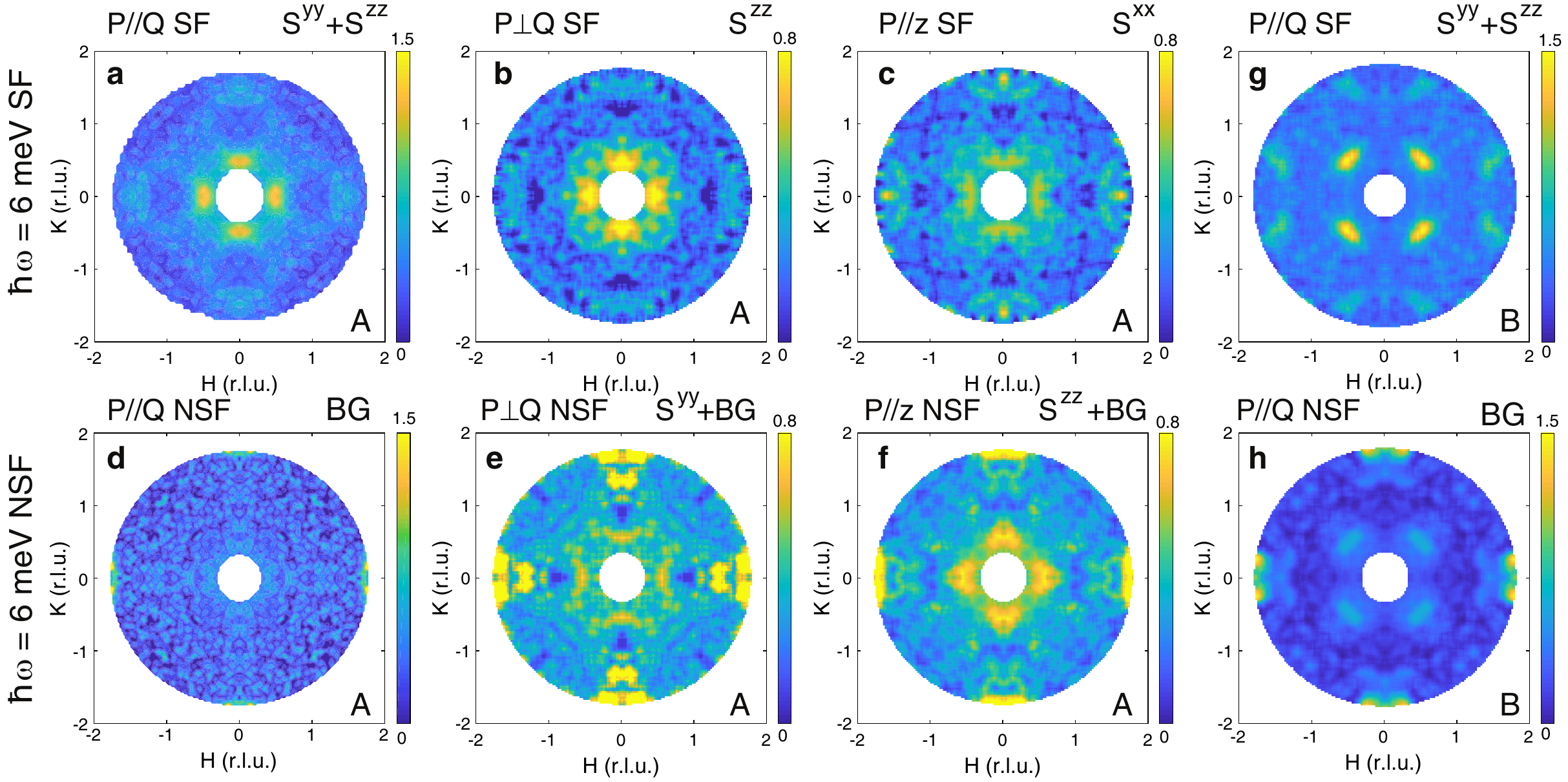}
\caption{{\bf Vector-polarized (XYZ) time-of-flight neutron spectroscopy of \FeTeSe\ Type A and Type B neutron samples.} (a-c) Pseudocolor plots for spin-flip (SF) channel with ${\bP} || {\bQ}$, ${\bP} \perp {\bQ}$, and ${\bP} || {\hat{z}}$ for sample A. (d-g) Pseudocolor plots for non-spin-flip (NSF) channel with ${\bP} || {\bQ}$, ${\bP} \perp {\bQ}$, and ${\bP} || {\hat{z}}$ for sample A. (g, h) Pseudocolor plots for spin-flip (SF) and non-spin-flip (NSF) with ${\bP} || {\bQ}$ for sample B. Note that the scattering patterns for sample A are different from the sample B, regardless of the initial neutron polarization. Data shown at energy transfer of 6(1) meV.}
\label{FigS2}
\end{figure}
\begin{figure}[p!]
\includegraphics[width=.7\textwidth]{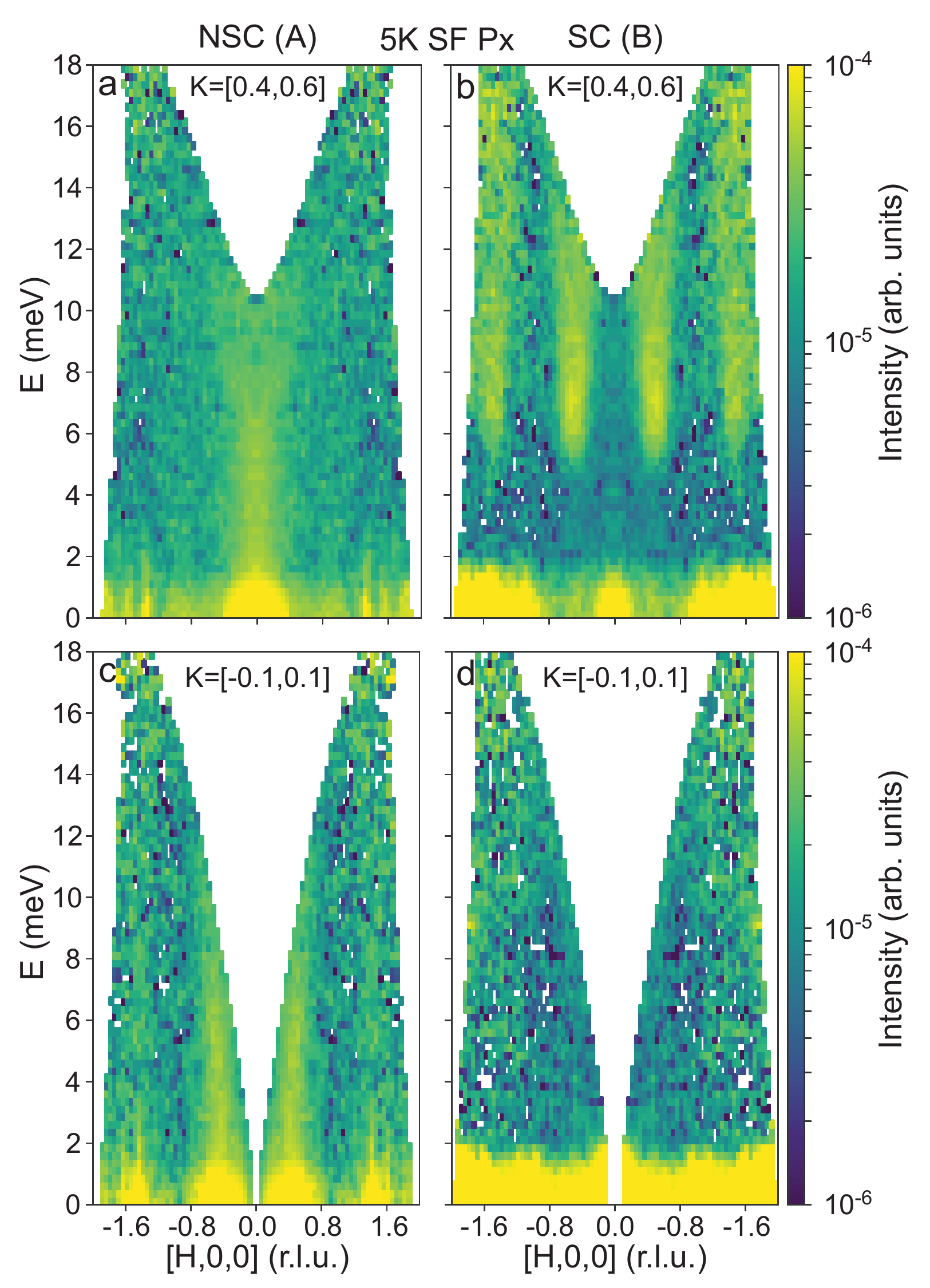}
\caption{{\bf Energy spectrum of double- and single-stripe magnetism of \FeTeSe\ Type A and Type B neutron samples.} (a, b) Pseudocolor plots of magnetic scattering intensity (SF channel with ${\bP} || {\bQ}$) in (H,E) plane at K = 0.5(1) for NSC sample A and SC sample B, respectively, at T = 5K. (c, d) The corresponding magnetic scattering intensity at T = 5K for samples A and B, respectively, at K = 0.0(1). Note, that in both cases the scattering is diffuse (broad) in energy. For NSC sample A the scattering is gapless and strongest near $(0,\pi)$ [(H, K) = (0, 0.5)]. For SC sample B the scattering is strongest near $(\pi,\pi)$ [(H, K) = (0.5, 0.5)] and has an energy gap of $\approx 5$~meV. All neutron data were symmetrized with respect to rotation and reflection symmetries in the $a-b$ plane.}
\label{FigS2.3}
\end{figure}
\begin{figure}[p!]
\includegraphics[width=1.\textwidth]{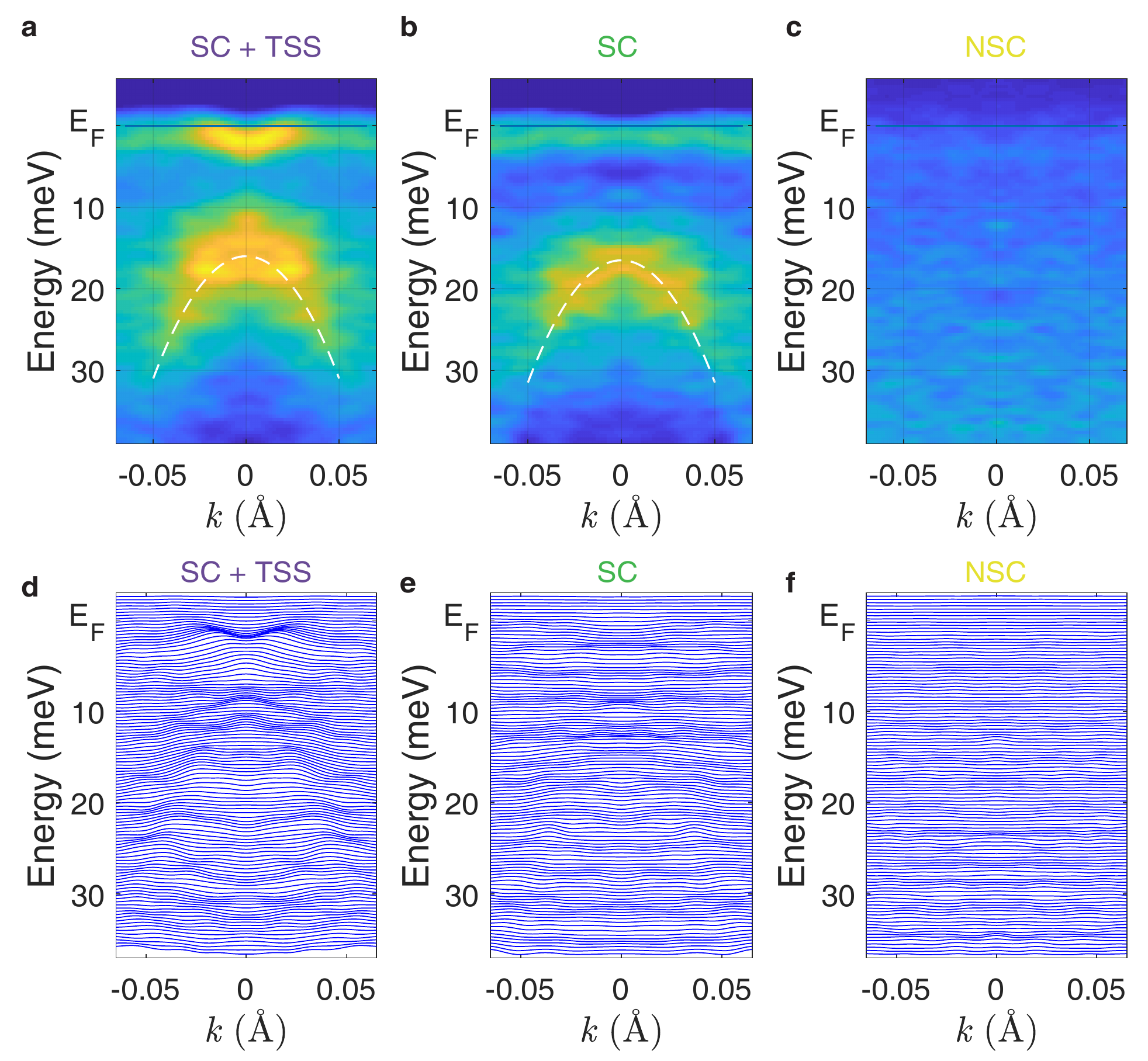}
\caption{{\bf Typical photoemission spectra for TSS+SC, SC, and NSC regions.} (a-c) Pseudocolor plots of symmetrized photoemission spectra (p-polarization) for TSS+SC, SC, and NSC regions, respectively. Blue line shows the chemical potential, $E_F$. White dashed lines indicate the bulk band. (d-e) Corresponding constant energy cuts (momentum distribution curves) of the photoemission spectra in (a-c), respectively.}
\label{FigS4}
\end{figure}
\begin{figure}[p!]
\includegraphics[width=1.\textwidth]{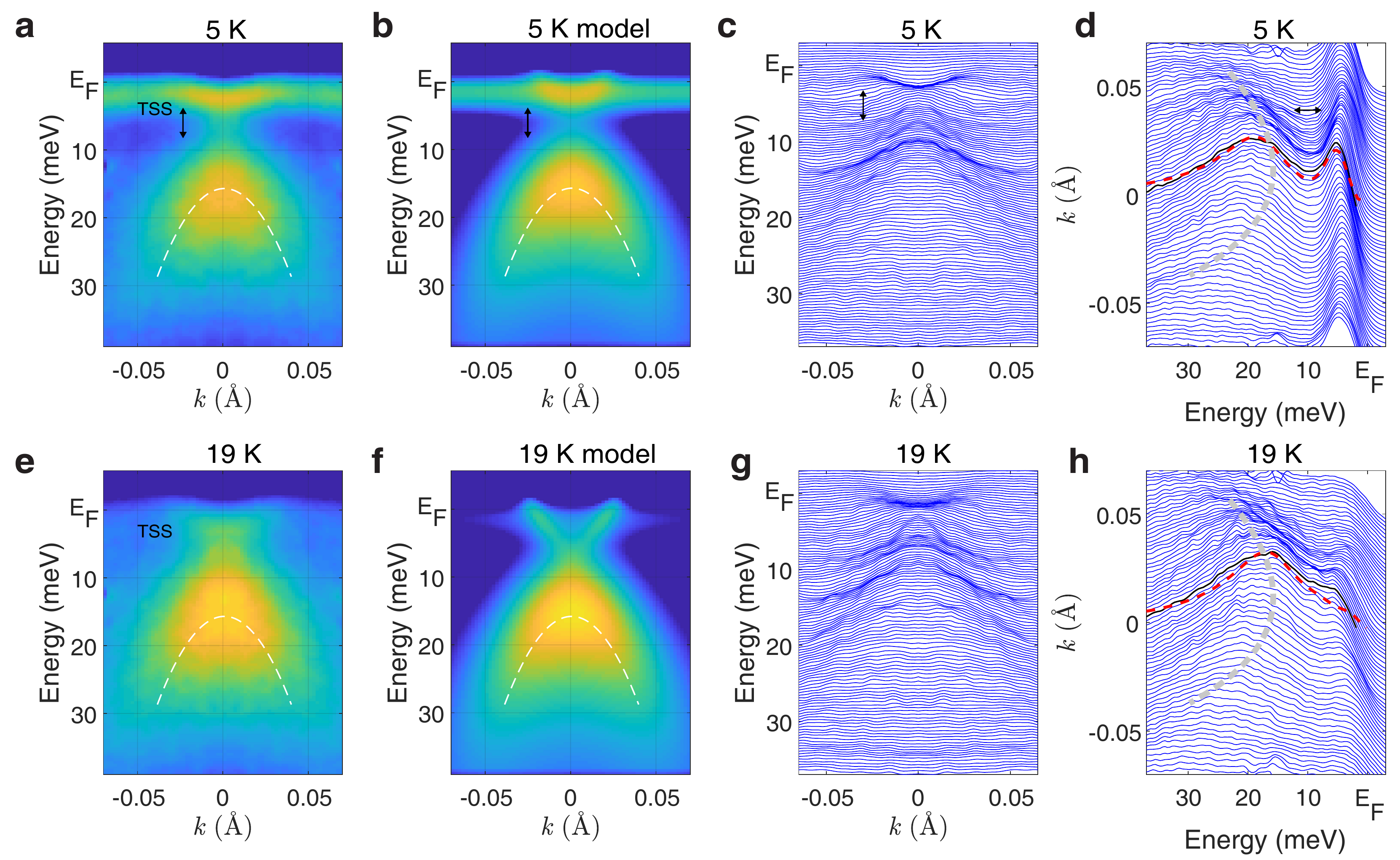}
\caption{{\bf Photoemission spectra for a typical TSS+SC region below and above T$_c$.} (a) Pseudocolor plot of symmetrized photoemission intensity (p-polarization) for TSS+SC region at 5 K. (b) Simulated photoemission intensity at 5 K using the three-component model, $I_\mathrm{B}+I_\mathrm{SC}+I_\mathrm{TSS}$, described in the text. (c, d) Waterfall plot of momentum distribution curves (MDCs) and energy distribution curves (EDCs), respectively, representing the 5 K data shown in (a). (e) Pseudocolor plot of symmetrized photoemission intensity (p-polarization) for the same TSS+SC region as in (a-d) at 19 K. (f) Simulated photoemission intensity at 19 K using the three-component model, $I_\mathrm{B}+I_\mathrm{SC}+I_\mathrm{TSS}$, described in the text. (g, h) Waterfall plot of MDCs and EDCs, respectively, representing the 19 K data shown in (e). To reduce the noise, data were collected with longer exposure time than typical scanning data shown in Fig.~\ref{FigS4}. Red dashed lines in (d) and (h) show the fitted EDCs at $k = 0$ calculated from the model shown in (b) and (f), respectively. White and grey dashed lines indicate the bulk band. Direct comparison of 5 K and 19 K data reveals opening of superconducting gap at $E_f$ and Dirac gap of TSS in SC state \cite{Rameau_2019,Zaki_2019}. }
\label{FigS4.1}
\end{figure}
\begin{figure}[p!]
\includegraphics[width=1.\textwidth]{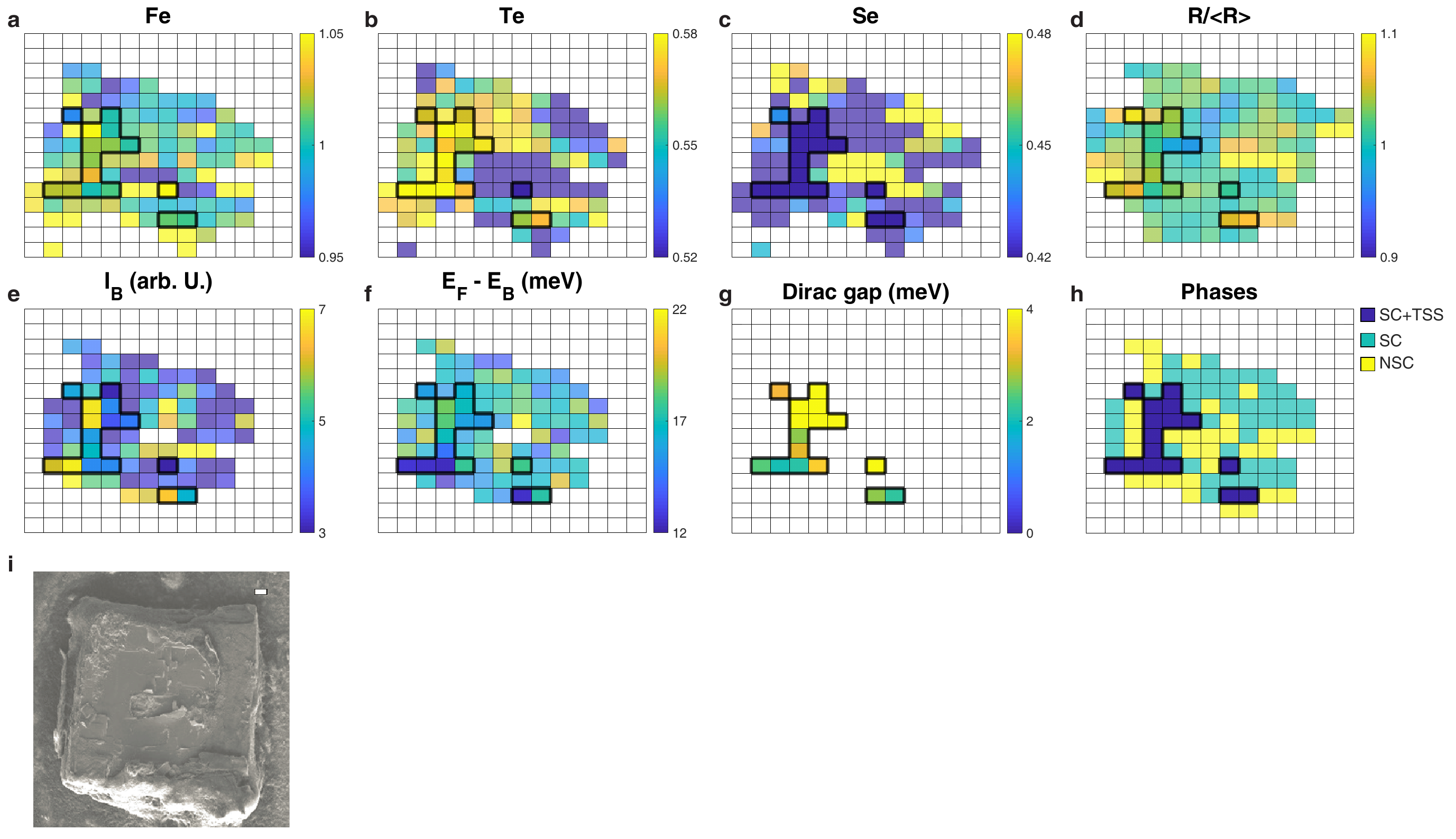}
\caption{{\bf Local chemical composition and electronic properties for additional \FeTeSe\ sample 1 (Type B).} (a-c) Distribution of Fe, Te, and Se concentrations. (d) Local resistance, $R$, at 300 K normalized to the average value, $\langle R\rangle$. (e-g) The integrated photoemission intensity of the bulk $d_{xz}$ band, $I_B$, bulk band gap, $E_F - E_B$, and massive Dirac gap of the topological surface state at 5 K. (h) Spatial distribution of the NSC, SC and SC+TSS phases in the sample identified from ARPES. The measured grid is $100 \mu$m$ \times 100 \mu$m. (i) SEM image of the cleaved crystal. White bar shows the 100 $\mu$m scale.}
\label{FigS5}
\end{figure}
\begin{figure}[p!]
\includegraphics[width=1.\textwidth]{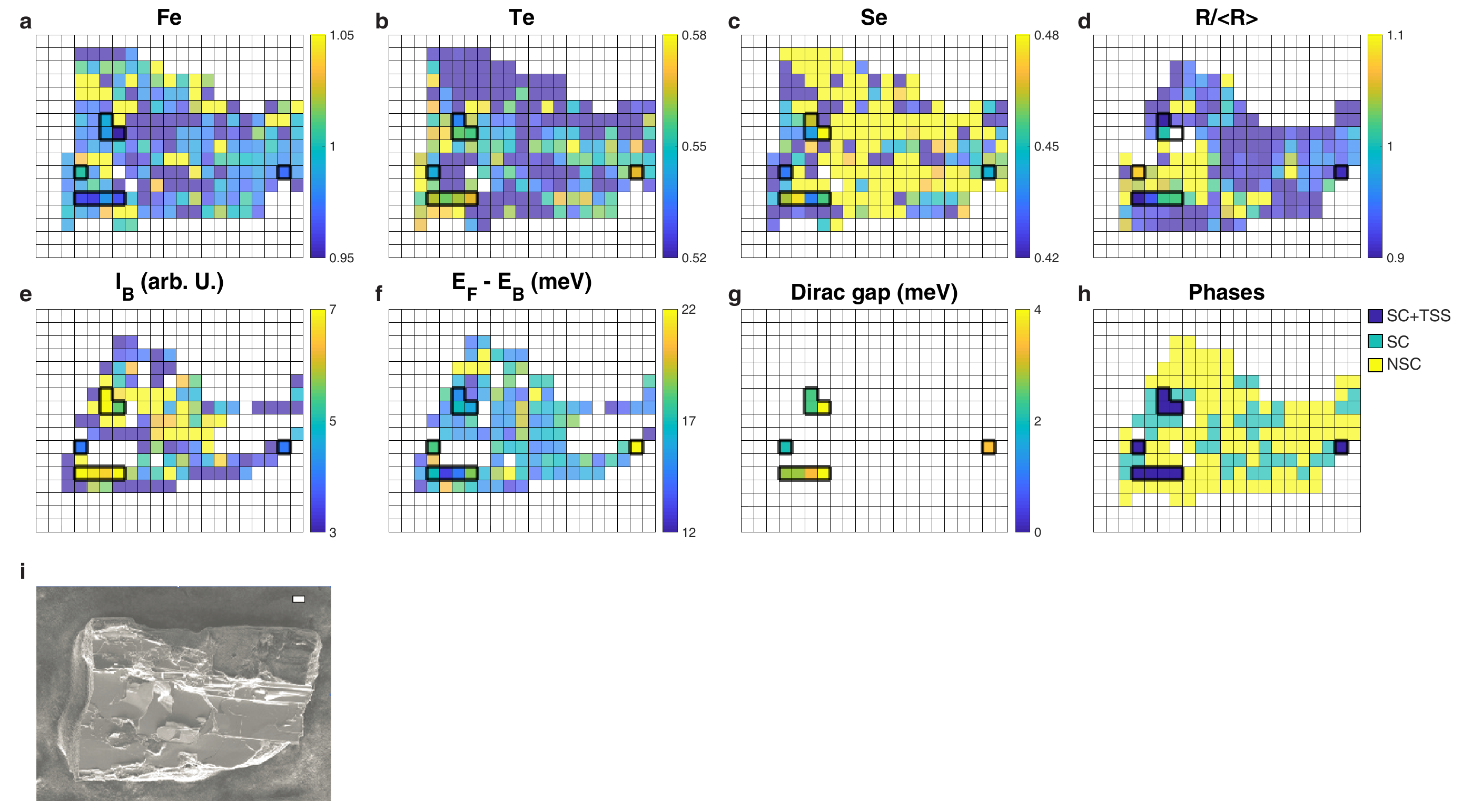}
\caption{{\bf Local chemical composition and electronic properties for additional \FeTeSe\ sample 2 (Type B).} (a-c) Distribution of Fe, Te, and Se concentrations. (d) Local resistance, $R$, at 300 K normalized to the average value, $\langle R\rangle$. (e-g) The integrated photoemission intensity of the bulk $d_{xz}$ band, $I_B$, bulk band gap, $E_F - E_B$, and massive Dirac gap of the topological surface state at 5 K. (h) Spatial distribution of the NSC, SC and SC+TSS phases in the sample identified from ARPES. The measured grid is $100 \mu$m$ \times 100 \mu$m. (i) SEM image of the cleaved crystal. White bar shows the 100 $\mu$m scale.}
\label{FigS6}
\end{figure}
\begin{figure}[p!]
\includegraphics[width=1.\textwidth]{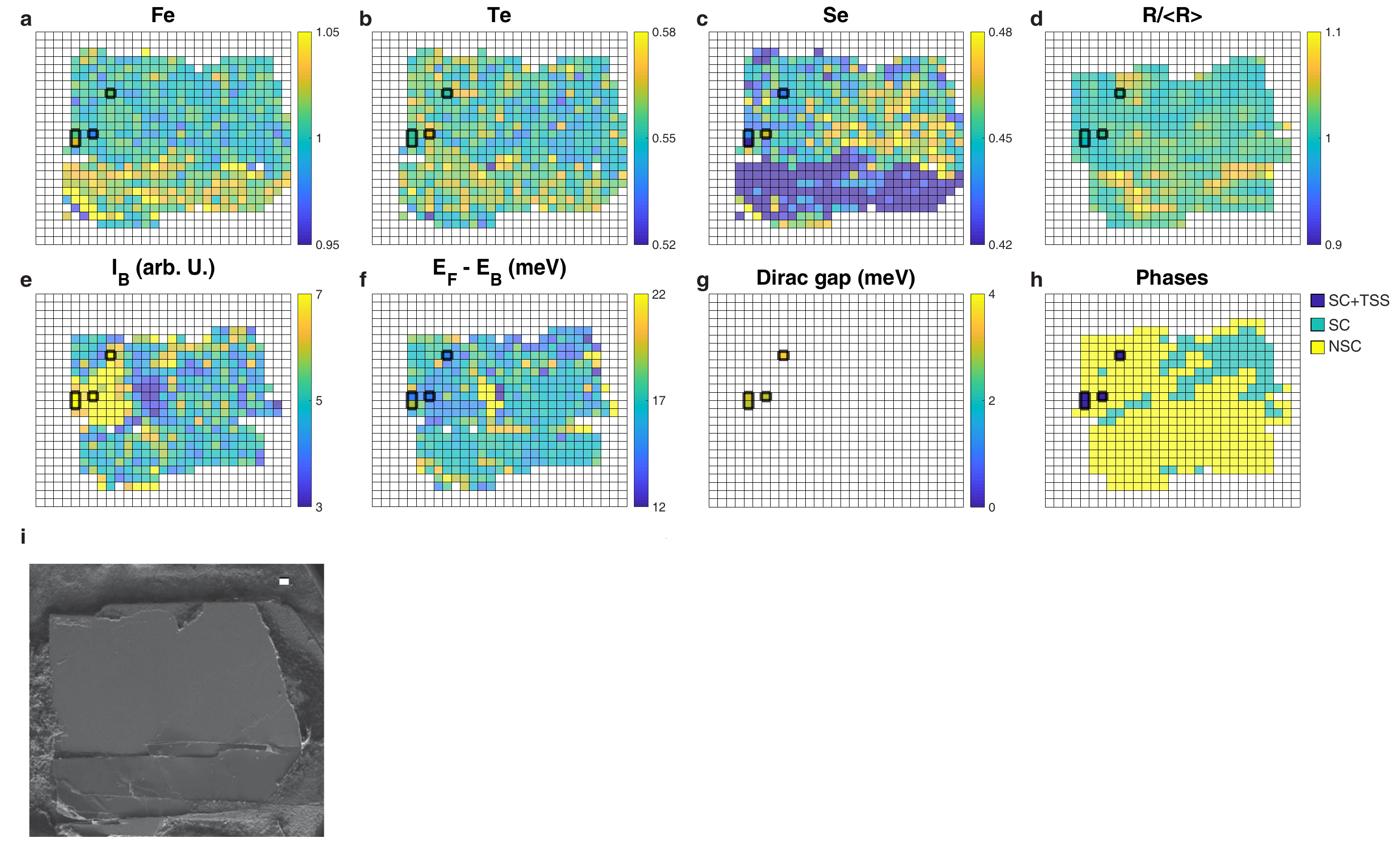}
\caption{{\bf Local chemical composition and electronic properties for additional \FeTeSe\ sample 9 (Type A).} (a-c) Distribution of Fe, Te, and Se concentrations. (d) Local resistance, $R$, at 300 K normalized to the average value, $\langle R\rangle$. (e-g) The integrated photoemission intensity of the bulk $d_{xz}$ band, $I_B$, bulk band gap, $E_F - E_B$, and massive Dirac gap of the topological surface state at 5 K. (h) Spatial distribution of the NSC, SC and SC+TSS phases in the sample identified from ARPES. The measured grid is $100 \mu$m$ \times 100 \mu$m. (i) SEM image of the cleaved crystal. White bar shows the 100 $\mu$m scale.}
\label{FigS7}
\end{figure}
\begin{figure}[p!]
\includegraphics[width=0.7\textwidth]{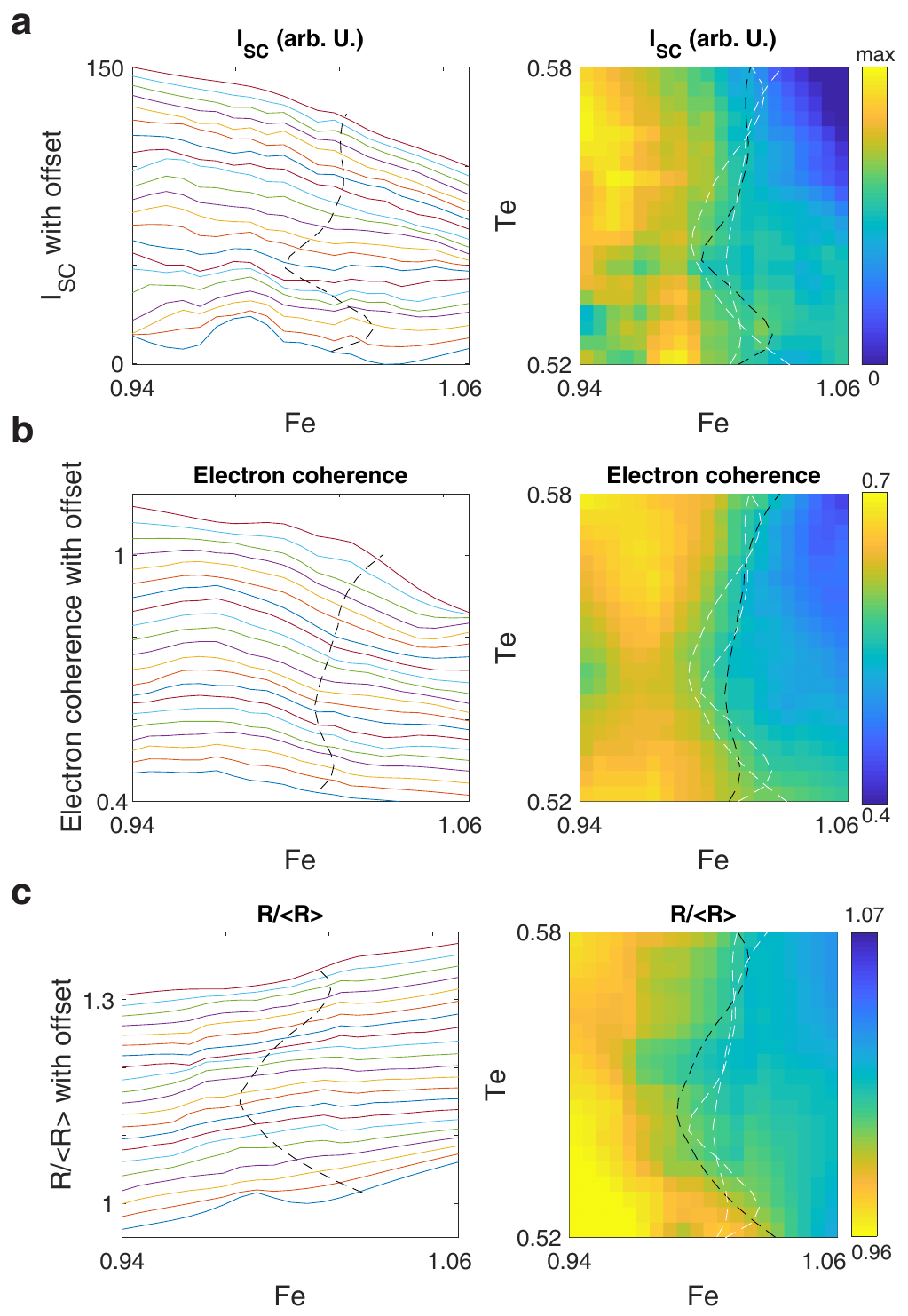}
\caption{{\bf SC-NSC phase boundary as a function of Fe content determined from different measurements.} (a) The line plots of $I_\mathrm{SC}$ as a function of Fe for different Te content (left panel) and the pseudocolor plot of $I_\mathrm{SC}$ as a function of Fe and Te (right panel). (b, c) The line (left) and pseudocolor (right) plots of electron coherence, $I_B/I_{tot}$, and $R/ \langle R \rangle$ and $R/ \langle R \rangle$ as a function of Fe and Te, respectively. Data were smoothed with the Savitzky-Golay algorithm, and the Fe phase boundary (black dashed curve) in each panel was determined based on an approximately 50 percent increase of the measured quantity shown in that panel. White dashed lines present the phase boundary as determined based on the measurements shown in the other two panels. The NSC-SC phase boundary shown in Fig.~\ref{Fig4:phase_diagram} of the main text is the average of the phase boundaries determined from different measurements. }
\label{FigS8}
\end{figure}

\end{document}